\documentclass[acmsmall, nonacm]{acmart}
\AtBeginDocument{%
  \providecommand\BibTeX{{%
    \normalfont B\kern-0.5em{\scshape i\kern-0.25em b}\kern-0.8em\TeX}}}

\setcopyright{acmcopyright}
\copyrightyear{2023}
\acmYear{2023}
\acmDOI{XXXXXXX.XXXXXXX}

\acmConference[CSCW '23]{The 26th ACM Conference On Computer-Supported Cooperative Work And Social Computing}{October 14--18,
  2023}{Minneapolis, MN}
\acmPrice{15.00}
\acmISBN{978-1-4503-XXXX-X/18/06}

\usepackage{amsmath}
\usepackage{pifont}
\newcommand{\cmark}{\ding{51}}%
\newcommand{\xmark}{\ding{55}}%
\usepackage{caption}
\usepackage{subcaption}

\usepackage{color-edits}
\addauthor{vc}{blue}
\addauthor{vera}{orange}
\addauthor{gb}{cyan}
\addauthor{JWV}{red}

\usepackage{booktabs}
\usepackage{multirow}

\begin{document}

\title[Understanding the Role of Human Intuition on Reliance]{Understanding the Role of Human Intuition on Reliance in Human-AI Decision-Making with Explanations}

\author{Valerie Chen}
\email{valeriechen@cmu.edu}
\affiliation{%
  \institution{Carnegie Mellon University}
  \country{USA}
}

\author{Q. Vera Liao}
\email{veraliao@microsoft.com}
\affiliation{%
  \institution{Microsoft Research}
  \country{Canada}
}

\author{Jennifer Wortman Vaughan}
\email{jenn@microsoft.com}
\affiliation{%
  \institution{Microsoft Research}
  \country{USA}
    }

\author{Gagan Bansal}
\email{gaganbansal@microsoft.com}
\affiliation{%
 \institution{Microsoft Research}
 \country{USA}
 }

\renewcommand{\shortauthors}{Chen, Liao, Vaughan, and Bansal}

\begin{abstract}
AI explanations are often mentioned as a way to improve human-AI decision-making, but empirical studies have not found consistent evidence of explanations' effectiveness and, on the contrary, suggest that they can increase overreliance when the AI system is wrong.  While many factors may affect reliance on AI support, one important factor is how decision-makers reconcile their own \emph{intuition}---beliefs or heuristics, based on prior knowledge, experience, or pattern recognition, used to make judgments---with the information provided by the AI system to determine when to override AI predictions.  We conduct a think-aloud, mixed-methods study with two explanation types (feature- and example-based) for two prediction tasks to explore how decision-makers' intuition affects their use of AI predictions and explanations, and ultimately their choice of when to rely on AI.  Our results identify three types of intuition involved in reasoning about AI predictions and explanations: intuition about the task outcome, features, and AI limitations. Building on these, we summarize three observed pathways for decision-makers to apply their own intuition and override AI predictions.  We use these pathways to explain why (1) the feature-based explanations we used did not improve participants' decision outcomes and increased their overreliance on AI, and (2) the example-based explanations we used improved decision-makers' performance over feature-based explanations and helped achieve complementary human-AI performance.  Overall, our work identifies directions for further development of AI decision-support systems and explanation methods that help decision-makers effectively apply their intuition to achieve appropriate reliance on AI.

\end{abstract}

\ccsdesc[500]{Human-centered computing~Collaborative and social computing}
\ccsdesc[500]{Computing methodologies~Artificial intelligence}

\keywords{Explainable AI, interpretability, human-AI interaction, decision support}

\received{20 February 2007}
\received[revised]{12 March 2009}
\received[accepted]{5 June 2009}

\maketitle

\section{Introduction}

Artificial intelligence (AI) systems---often based on machine learning (ML) models---are increasingly used to support decision-makers, even in high-stakes domains like healthcare and finance. Given the complexity of AI systems, it is often suggested that decision-makers could benefit from access to explanations of their predictions.  The hope is that such explanations will help decision-makers reason about when and when not to rely on the AI system's predictions, achieving {\em appropriate reliance}~\cite{bansal2021does}.
However, across many domains, empirical studies of explanations have produced mixed  results~\cite{lai2019human,liu2021understanding,bansal2021does,zhang2020effect,poursabzi2021manipulating,wang2021explanations, buccinca2020proxy,kim2022hive}, with some suggesting that explanations 
increase decision-makers' tendency to rely on AI predictions even when the AI system is wrong~\cite{zhang2020effect,poursabzi2021manipulating,bansal2021does,wang2021explanations}---a phenomenon often referred to as \emph{overreliance}~\cite{buccinca2021trust,vasconcelos2022explanations}. 

To design explanations that better support appropriate reliance, we must first understand the process through which decision-makers determine whether to rely on AI. Prior studies have typically focused on decision outcomes alone by measuring aggregate performance rather than studying decision-making processes~\cite{lai2021towards},
limiting our ability to make generalizable recommendations on how and when explanations can help. One particular blind spot in the existing literature is around the role of a decision-maker's own \emph{intuition} and how it is integrated with AI predictions and explanations during the decision-making process. 

While various definitions of ``intuition'' exist~\cite{gigerenzer2007gut,rosenblatt1994intuition,shirley1996intuition}, we follow recent work on human-AI decision-making~\cite{chen2022machine,potanvcok2019role} and use the term to broadly refer to beliefs or heuristics that people bring into the decision based on their domain knowledge, experience, instinct, or pattern recognition.
For example, a radiologist looking at an X-ray knows where to look and what to look for, and may form a quick opinion about the diagnosis. When given an AI system to support their diagnosis, this intuition may be in agreement or disagreement with the information provided by the AI system.
Prior work has suggested that decision-makers' confidence in their own intuition affects reliance on AI~\cite{lu2021human} and that their intuition about what features matter to a decision can facilitate the detection of model errors from AI explanations~\cite{chen2022machine}. However, little is known about the process through which this happens and how it depends on the type of explanation used.

In this work, we take a bottom-up approach, using a think-aloud protocol to investigate participants' decision-making process with AI predictions and explanations.
Our goal is to more holistically understand what types of intuition are involved in engaging with AI decision-support systems and the interplay between these types of intuition and different types of explanations. In light of the well-known pitfall that AI explanations can increase overreliance~\cite{zhang2020effect,bansal2021does,wang2021explanations,poursabzi2021manipulating}, we take a particular interest in decision-makers' paths to \emph{non-reliance} (i.e., overriding the AI prediction) when the AI system is incorrect. Specifically, we conduct a mixed-methods study (N=26) to investigate how participants make decisions with two common types of explanations: feature- and example-based. Feature-based explanations consist of scores or weights that describe the extent to which each feature of a decision instance contributed to the model's prediction.  Example-based explanations support case-based reasoning by providing examples of similar instances and the model's prediction (and often also their ground truth labels~\cite{lai2021towards}). We study these explanations in the context of two decision tasks based on different types of data (tabular and text).

Our exploratory quantitative analysis shows that, consistent with prior work \cite{zhang2020effect,bansal2021does,wang2021explanations}, feature-based explanations did not improve participants' decision outcomes compared to their performance without AI support, and led to overreliance on incorrect AI predictions. In contrast, example-based explanations reduced overreliance and helped achieve human-AI \emph{complementary performance}---outperforming human or AI alone. We analyze participants' think-aloud data during the decision tasks to explain these observations. In summary, our work makes the following contributions:

\begin{itemize}
    \item \textit{Contributing to a fundamental understanding of people's decision-making process with AI predictions and explanations:} We leverage a think-aloud protocol and mixed-methods study to investigate this process.
    Our analysis highlights three types of intuition that decision-makers apply to override the AI prediction when they believe it is wrong, henceforth referred to as {\em intuition-driven pathways}: (1) strong intuition about the decision \textit{outcome} that disagrees with the AI prediction; (2) intuition about \textit{features} which they use to reason about explanations and identify evidence that discredits the AI prediction; and (3) intuition about \textit{AI limitations} that they use to infer \textit{prediction unreliability}. 
    \item \textit{Demonstrating and explaining the effectiveness of example-based explanations for decision support:}
    Our experiment looks beyond the commonly studied feature-based explanations. We find them less effective in supporting decision-making than an example-based explanation that presents the AI prediction and ground-truth labels for two similar training examples. Using the set of intuition-driven pathways, we identify the benefits of example-based explanations, including less disruption of and more compatibility with people's natural intuition formation process, supporting inductive reasoning with additional context to form new intuition about features, and appropriately signaling prediction unreliability. 
    \item \textit{Design implications for AI decision-support systems:} Based on these findings and participants' post-study interviews, we suggest design recommendations for explainable AI methods and AI decision-support tools more broadly that better accommodate these intuition-driven pathways to support appropriate reliance on AI.
\end{itemize}{}

\section{Related Work and Research Questions}

First, we overview related work on explainable AI (XAI), focusing on XAI for ML models---what is sometimes referred to as \emph{interpretability} in the ML literature~\cite{lipton2018mythos,doshi2017towards}. 
We then overview why XAI is believed to be useful for human-AI decision-making and gaps in the community's understanding. 

\subsection{Overview of XAI}\label{related:xai}

Given the increasing use of AI and ML systems, there is a growing need for people interacting with these systems to understand the underlying models. The technical field of explainable AI (XAI) grew out of these concerns. A diverse set of XAI methods have been proposed that surface technical details of ML models, as surveyed by several authors~\cite{adadi2018peeking,arrieta2020explainable,gilpin2018explaining,carvalho2019machine,guidotti2018survey}. XAI methods can be broadly grouped into two categories: inherently interpretable models that are thought to be intuitive to understand (e.g., rule-based models and linear regressions), and post-hoc techniques that generate explanations for complex models like deep neural networks and random forests.
Explanations can also be classified as global or local~\cite{adadi2018peeking, guidotti2018survey}. Global explanations summarize a model's overall behavior, while local explanations shed light on the model's behavior for a particular instance.  

Our study focuses on two types of local, post-hoc explanations: feature-based and example-based.
Feature-based explanations assign a value to each feature that is generally interpreted as its  ``contribution'' to a given prediction. Popular algorithms to generate feature-based explanations include LIME~\cite{ribeiro2016should}, SHAP~\cite{lundberg2017unified}, GradCAM~\cite{selvaraju2017grad}, and Integrated Gradients~\cite{sundararajan2017axiomatic}. 
Example-based explanations typically select ``representative samples'' from the training set, with two common approaches to selecting these samples. One is to select prototypes that are representative of a prediction class 
to help people understand why the model predicts the current instance belongs to that class~\cite{kim2022hive,hase2020evaluating,buccinca2020proxy,cai2019effects}. The other is to select examples that resemble the current instance and show the AI predictions (and ground truth) on those examples. This approach helps people understand not only how the model makes decisions but also how it might make mistakes~\cite{lai2019human, binns2018s,dodge2019explaining}.

Despite the rapid development of XAI techniques, there remain open questions about what these methods are useful for~\cite{lipton2018mythos, chen2022interpretable}. Additionally, researchers in the AI, HCI, and CSCW communities have called for more human-centered approaches~\cite{vaughan2021humancentered,liao2021human,ehsan2020human,wang2019designing} to investigate what people need and how they interact with AI explanations in specific use cases. Common use cases of AI explanations include supporting model debugging, assisting decision-making, auditing models, and knowledge discovery~\cite{arrieta2020explainable,chen2022machine,liao2022connecting,suresh2021beyond}. Aligning with this line of work, we study how people interact with explanations, focusing on the use case of assisting decision-making.

\subsection{Human-AI Decision-Making}

Human-AI decision-making, also referred to as AI-assisted decision-making, broadly encompasses set-ups where an ML model is used to help users to make a final judgment or decision~\cite{lai2021towards,green2019disparate,de2020case,erlei2020impact}---often considered as a form of collaboration between human and AI system.
While AI assistance typically provides an ML model's predictions, there is growing work studying whether additional information---performance measures~\cite{yin2019understanding}, explanations~\cite{bansal2021does,wang2021explanations}, or information about prediction uncertainty~\cite{rechkemmer2022confidence,zhang2020effect}, for example---can further improve decision outcomes~\citep{lai2021towards}. 
Among other goals, a common interest is to study what form of AI assistance can help decision-makers outperform both human and AI alone to achieve human-AI \textit{complementary performance}~\cite{liu2021understanding,zhang2020effect}. 
However, many empirical studies on human-AI decision-making did not observe complementary performance~\cite{lundberg2018explainable, green2019principles, beede2020human, carton2020feature, zhang2020effect, lai2020chicago, yang2020visual, liu2021understanding, poursabzi2021manipulating,wang2021explanations}.
To achieve complementary performance, it is important to encourage \textit{appropriate reliance}~\cite{zhang2020effect,bansal2021does,lu2021human,wang2021explanations,cao2022understanding}---following the AI system when it is likely to be correct and not following it when it is likely to be wrong.

\subsubsection{Can XAI methods improve appropriate reliance?} Prior work has asked whether providing decision-makers with explanations of an AI system's predictions can improve appropriate reliance, since decision-makers might be less likely to follow an AI prediction if an explanation suggests flawed model reasoning~\cite{bayati2014data,bussone2015role,caruana2015intelligible}. As surveyed by~\citet{lai2021towards}, the majority of prior empirical studies of AI decision support have focused on feature-based explanations, with a limited set of studies on other explanation types like example-based explanations, rule-based explanations, and counterfactual explanations. 
These studies have been conducted across various decision tasks with different data types including text data~\cite{lai2019human,liu2021understanding,bansal2021does,he2023knowing}, tabular data~\cite{zhang2020effect,poursabzi2021manipulating,wang2021explanations}, and image data~\cite{buccinca2020proxy,kim2022hive}.

Unfortunately, empirical studies have generally failed to confirm that providing explanations can improve appropriate reliance. On the contrary, several studies found that showing feature-based explanations increases people's tendency to \emph{over-rely} on the model when it is wrong, compared to showing only the AI predictions~\cite{zhang2020effect,bansal2021does,poursabzi2021manipulating,wang2021explanations}.
\citet{wang2021explanations} found that example-based explanations underperform feature-based explanations for improving appropriate reliance, though other studies suggest example-based explanations are better at helping people detect model errors~\cite{buccinca2020proxy,cai2019effects,kim2022hive}, usually by recognizing the dissimilarity between the instance and selected examples. However, we note that these studies utilized different kinds of example-based explanations and investigated different tasks with different types of data (tabular, text, and image). Our work aims to understand the underlying causes of overreliance with different kinds of explanations. We return to these studies to interpret their mixed results in the context of our findings in Section~\ref{sec:explainpriorwork}.

Initial efforts to explain why AI explanations increase overreliance have focused on the role of cognitive engagement. The hypothesis is that participants (often recruited on crowdsourcing platforms) over-rely on the AI system because they do not deeply engage with the explanations. 
The dual processing model has been cited as a useful framework~\cite{kaur2020interpreting,buccinca2020proxy,gajos2022people,liao2021human,liao2022designing}: instead of engaging in analytical reasoning with the explanations (system 2 thinking), people may invoke heuristics to make a quick judgment (system 1 thinking), including the judgment of ``just deferring to AI.''\footnote{Note the two systems reflect the depth of reasoning instead of content. Heuristics can also be used as part of system 2 thinking~\cite{chen1999heuristic,petty2011elaboration}. While our investigation of intuition includes heuristics based on domain knowledge (which differs from a superficial heuristic to defer to AI), we do not claim whether participants engaged \emph{only} in system 1 or system 2 thinking. However, the think-aloud setup might have forced participants to be relatively more engaged.} 
A recent study~\cite{ehsan2021explainable} also suggested that people often invoke positive heuristics that superficially associate AI being explainable with it being trustworthy, which can lead to overreliance. 
~\citet{buccinca2021trust} showed that incorporating cognitive forcing functions, which aim at deepening cognitive engagement (e.g., slowing people down), can improve appropriate reliance with explanations, but at the cost of worsened subjective experience from more effortful interactions. A recent work by~\citet{vasconcelos2022explanations} further elucidates this lack of cognitive engagement through a cost-benefit framework, which shows that people strategically choose to engage cognitively with explanations by weighing the costs of engaging against simply deferring to the AI system. This work suggests that a fundamental issue of current XAI techniques is that they are too effortful to verify, thus discouraging cognitive engagement.

\subsubsection{What is the role of human intuition on reliance?}
\label{sec:intuition} Even if a decision-maker is motivated to engage with explanations, \textit{how} do they decide to rely or not rely on the AI system? This is the overarching question that motivates our study. Most relevant to our study is a theoretical work by~\citet{chen2022machine}, which highlights the role of \textit{human intuition} in reasoning about explanations. They propose a conceptual framework that distinguishes between model decision boundaries (how the model makes decisions using features), and task decision boundaries  (how the decision \textit{should} be made). Their framework points out that popular feature-based explanations only facilitate understanding of the former. 
Yet, in order to detect model errors, decision-makers must apply their (correct) intuition about the task decision boundaries and contrast them with the model decision boundaries revealed by the explanation.
The authors suggest two types of human intuition useful for detecting model errors---intuition about feature \textit{relevance} and intuition about feature \textit{mechanism}
(e.g., weight), which align with prior findings about how experts critique model explanations in interactive ML settings~\cite{ghai2020explainable,stumpf2009interacting}. To our knowledge, beyond this theoretical work, there have not been empirical investigations into what types of human intuition are involved in human-AI decision-making.

While we follow~\citet{chen2022machine} and use the term ``intuition'' to broadly refer to beliefs and heuristics people bring to the decision, we note that many definitions of intuition exist in different disciplines~\cite{gigerenzer2007gut,rosenblatt1994intuition}. One hallmark of intuition is the reliance on knowledge and beliefs stored inside one's cognition, rather than applying formal reasoning to a complete set of information. Intuition has therefore been widely studied for expert decision-making~\cite{salas2010expertise} and decision-making under uncertainty~\cite{hall2002reviewing}. While intuition is often used interchangeably with terms like heuristics, insights, and instinct, literature surveys on this topic acknowledge that there are no agreed-upon sources of intuition~\cite{shirley1996intuition,salas2010expertise}. Indeed, sources span domain expertise, experience, associative memory, pattern recognition, emotional and affective awareness, and more. In this work, we focus on investigating \emph{what} types of intuition participants bring to make decisions, without drawing conclusions on \textit{how} they are generated (which is not permitted by our think-aloud method).

A source of human intuition that has been studied in the context of human-AI decision-making is one's domain expertise about the decision task.  
One line of work explored whether and how decision-makers with more domain expertise may benefit from AI decision support differently. For example, people with higher overall performance on their own (a proxy for more domain knowledge) tend to achieve higher performance when AI is introduced~\cite{inkpen2022advancing,liu2021understanding}. Higher confidence in one's own decisions compared to novices could also impact interactions with AI support~\cite{lu2021human}. For example, ~\citet{cheng2022heterogeneity} found that less experienced child welfare caseworkers tend to make decisions more closely following algorithmic risk scores, while senior workers tend to engage in further screening on their own.

\subsection{Research Questions}

In summary, our work contributes to the growing area of research on XAI for decision support by studying the role of human intuition when decision-makers interact with such systems. To understand why prior studies have found that XAI support is ineffective, and even increases overreliance, we move beyond quantitatively studying decision outcomes to investigating the decision-making process. Different from qualitative studies that aim to understand the experience of decision-makers retrospectively~\cite{kawakami2022improving,cai2019human,park2021human,jacobs2021designing,wang2021brilliant}, we adopt a think-aloud protocol to investigate the real-time process as participants engage with the AI output~\cite{hoffman2018metrics,zhang2022towards}, and use the data to understand why two types of explanations---feature-based and example-based---improve or inhibit appropriate reliance. 
We focus on the following research questions: 

\begin{itemize}
    \item \textbf{RQ1:} What types of human intuition are involved in engaging with AI predictions and explanations, and how do they affect reliance on AI? 
    \item \textbf{RQ2:} Does human intuition come into play differently with feature- and example-based explanations, and do these differences explain their different effects (if any) on decision accuracy and appropriate reliance? 
\end{itemize}

In addition, we explore participants' subjective experiences with the two types of AI explanations. Our aim is to inform the future development of XAI techniques and broader approaches to provide better AI support for decision-making.

\section{Methods}

We describe the set-up of our study and analysis, the two prediction tasks that we asked participants to engage with, and the types of explanations they were shown. We then overview our experimental design and study procedure and discuss the approaches used to analyze the data collected during our study.  We note that participation was voluntary and the study was IRB approved.

\subsection{Participants}

Since our research questions do not target any specific population, we chose to start with a convenience sampling strategy and then diversify our selection from a large pool of sign-ups based on their background.
Specifically, we advertised our study on Twitter and internal message boards within a large, international technology company to target a variety of communities including researchers, ML engineers, and cross-functional teams. We limited participation to people located in the US since one of our tasks required participants to make judgments about US salaries. 
In the sign-up form, we inquired about education level, job role, and self-reported knowledge about ML and XAI.  
237 people signed up, and we selected 26 participants by diversifying on ML and XAI backgrounds.

Our participants were from both industry and academia. 
73\% had graduate degrees, while the remaining 27\% had college degrees. 
The common roles participants held were software engineer, data scientist, research engineer/scientist, and Ph.D. student.
In terms of self-reported experience with ML or AI tools, 4\% had no experience, 23\% had limited experience, 31\% had used them often, and 42\% consider themselves experts.
In terms of self-reported experience with XAI, 15\% had no experience, 15\% had limited experience, 54\% had used them often, and 16\% considered themselves experts. Appendix~\ref{sec:participanttable} contains more details on participants. 
Despite our effort to diversify, our participants were skewed toward an ML-experienced and highly educated population. We acknowledge that this is a potential limitation of our study. However, we note that our primary focus is a qualitative understanding of the role of intuition, rather than quantifying its effect or distribution. 
Furthermore, during exploratory analysis, we tested the effect of participants' background attributes gathered in the sign-up form and did not find any significant effect. Thus, we proceeded with analyzing all participants' data together.

\subsection{Prediction Tasks}

\begin{figure}[t!]
     \centering
     \begin{subfigure}[b]{0.48\textwidth}
         \centering
         \includegraphics[width=0.99\textwidth]{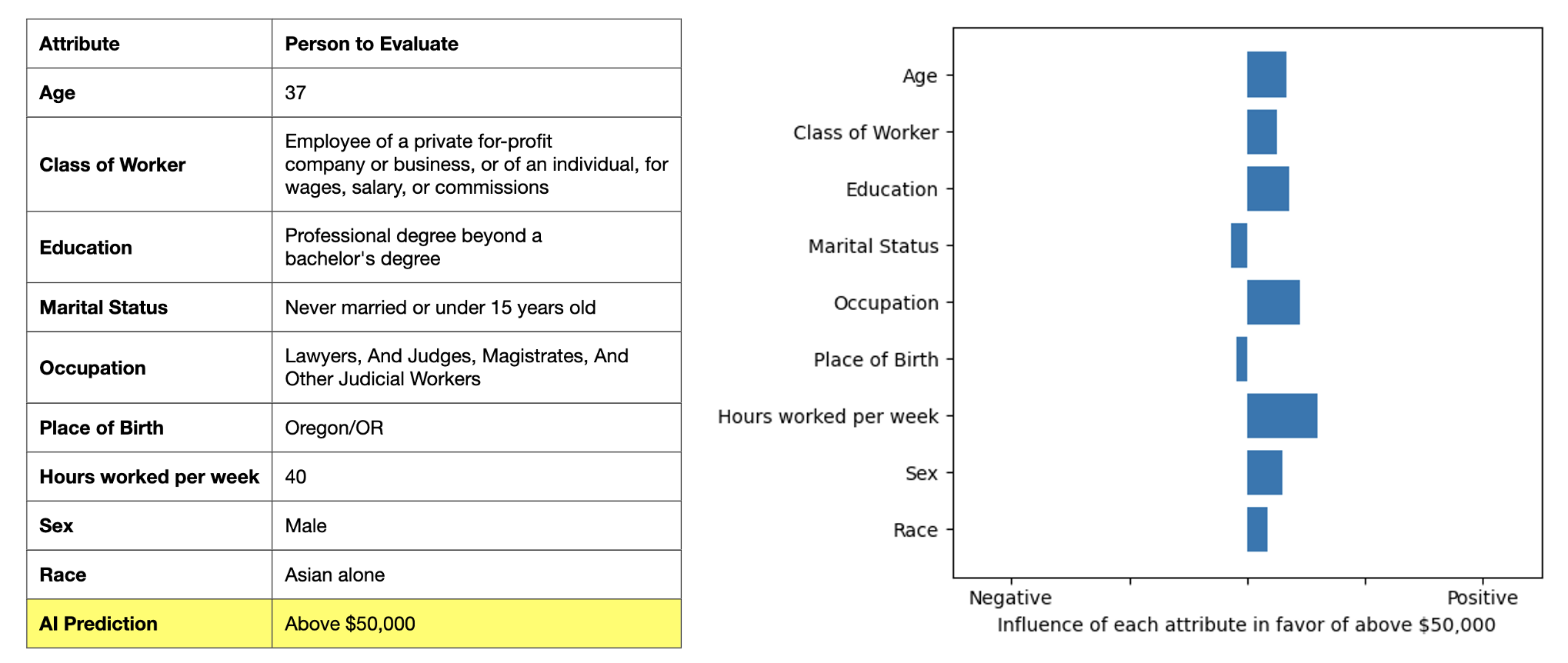}
         \caption{Feature-based explanation for income prediction.}
         \label{fig:tab_part1}
     \end{subfigure}
     \hfill
     \begin{subfigure}[b]{0.48\textwidth}
         \centering
         \includegraphics[width=0.99\textwidth]{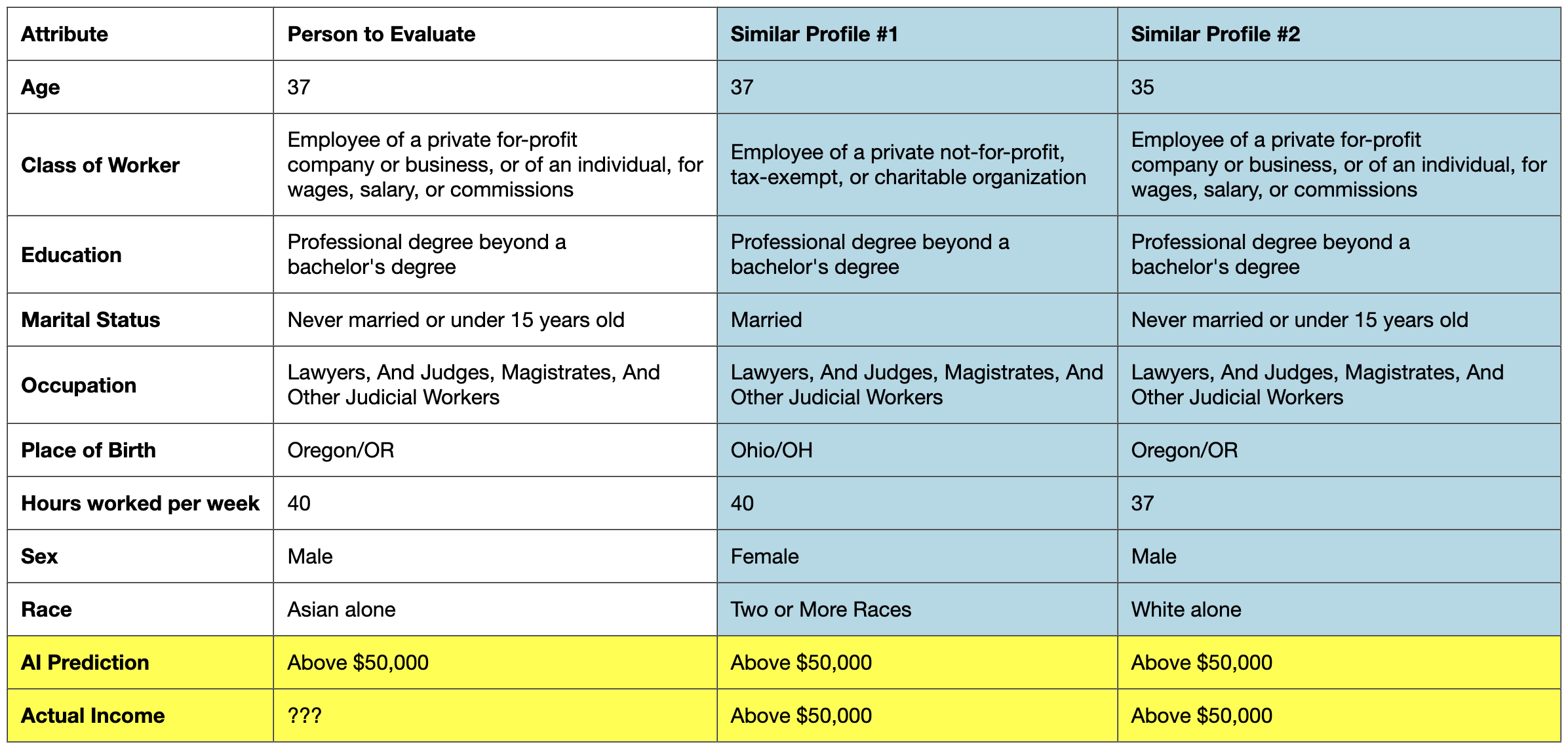}
         \caption{Example-based explanation for income prediction.}
         \label{fig:tab_part2}
     \end{subfigure}
          \centering
     \begin{subfigure}[b]{0.48\textwidth}
         \centering
         \includegraphics[width=0.99\textwidth]{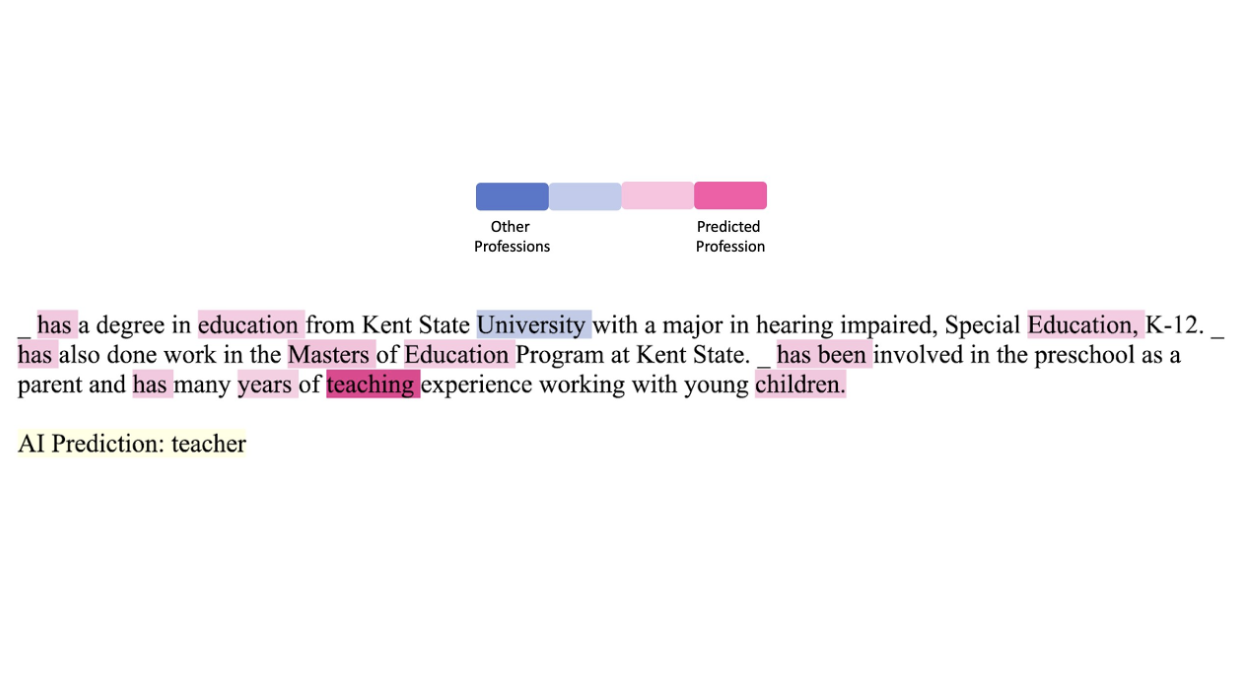}
         \caption{Feature-based explanation for biography classification.}
         \label{fig:text_part1}
     \end{subfigure}
     \hfill
     \begin{subfigure}[b]{0.48\textwidth}
         \centering
         \includegraphics[width=0.99\textwidth]{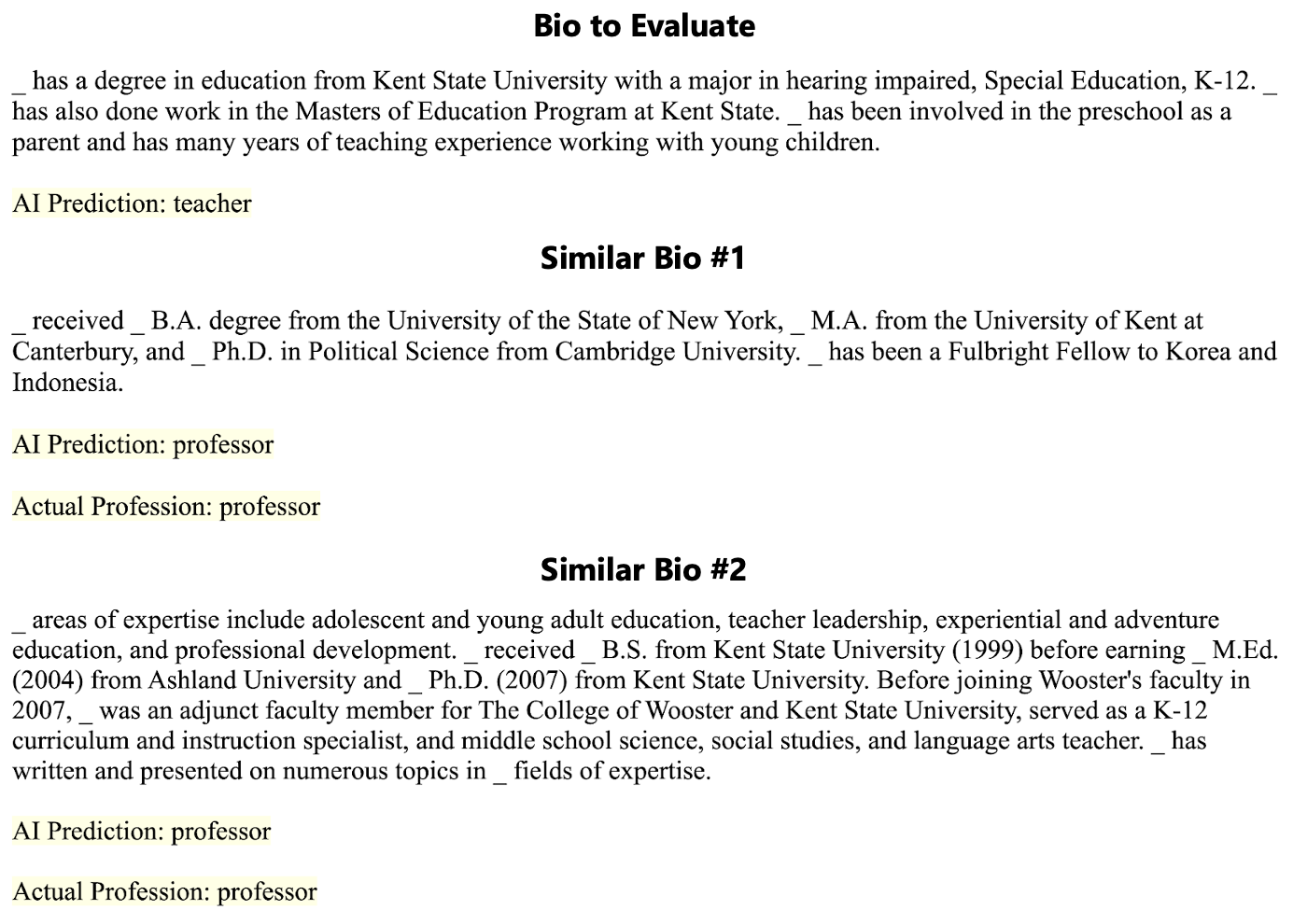}
         \caption{Example-based explanation for biography classification.}
         \label{fig:text_part2}
     \end{subfigure}
     \caption{Examples of the feature- and example-based explanations (columns) for each of the two prediction tasks of income prediction and biography classification (rows).}
     \label{fig:example_explanations}
\end{figure}

To select two prediction tasks for the user study, we considered the following four criteria: (1) the tasks should not require specialized expertise, but rather the general population should be able to reason about and perform the tasks; (2) the tasks should be difficult enough that participants cannot obtain high accuracy without AI assistance; (3) each individual decision should take no more than a couple of minutes to make; and (4) the selected tasks should have different data modalities (e.g., tabular and text) so that we can better understand the generalizability of our results. Using these criteria, we selected two tasks: income prediction and biography classification. Figure~\ref{fig:example_explanations} shows examples of both tasks.

\paragraph{Income Prediction.} In this task, the participant judges whether an individual's income is more than \$50,000 USD given a profile that includes the individual's age, class of work (e.g., employed at a private, for-profit company), education level, marital status, occupation, place of birth, number of hours worked per week, sex, and race. The cut-off of \$50,000 is close to the median income in the US at the time the data was generated.
The AI system used in this task is a random forest model that we trained using data from the Folktables dataset~\cite{ding2021retiring} of individuals sampled from the 2018 US Census data. This model achieved an accuracy of 80.8\% on a held-out test set.
As shown in Figure~\ref{fig:example_explanations}~(a), participants were given an individual's profile displayed in a table with the AI prediction (when applicable) highlighted at the bottom.

\paragraph{Biography classification.} In this task, the participant is asked to guess an individual's profession based on the individual's online biography. The AI system used in this task was trained on the BIOS dataset~\cite{de2019bias}, which contains hundreds of thousands of online biographies from the CommonCrawl corpus. While the original dataset consisted of individuals with 29 professions, we simplified the task by narrowing it down to five professions, following Liu et al.~\cite{liu2021understanding}: psychologist, physician, surgeon, teacher, and professor. We embedded each biography using a bag-of-words approach and trained a random forest model on this embedding. This model achieved an accuracy of 75.4\% on a held-out test set.   As shown in Figure~\ref{fig:example_explanations} (c), participants were given the full text of the individual's biography with the AI prediction (when applicable) displayed below the text.

\subsection{Explanation Methods} \label{sec:exp_types} 

Participants were shown two types of explanations: feature-based and example-based.

\paragraph{Feature-based explanations} 
We generated feature-based explanations using LIME~\cite{ribeiro2016should}, a post-hoc explanation technique.\footnote{Specifically, we used the open-source implementation of LIME: \url{https://github.com/marcotcr/lime}}
LIME works by computing a simple linear approximation of the model's decision boundary in the local region of the instance to be explained and using the coefficients as a measure of each feature's contribution to the model's prediction. 
We chose LIME because it is widely used and can be readily applied to both tasks (with tabular and text data).
For the income prediction task, we presented the coefficients in a bar chart as in Figure~\ref{fig:example_explanations} (a). Following~\citet{hadash2022improving}, we fix the direction of feature contribution to mean ``increasing income level'' so features with positive coefficients contributing to a prediction of ``above \$50,000'' have bars on the right and features with negative coefficients to such a prediction have bars to the left. The length of the bar represents the magnitude of the contribution.
For the biography classification task, we used a standard approach~\cite{madsen2021post} of color-highlighting the words (features), particularly those with the top 10 highest-magnitude coefficients, as shown in Figure~\ref{fig:example_explanations} (c). 
Words with positive coefficients (indicating that they support the model's prediction) are highlighted in red, while words with negative coefficients (indicating that they support other professions) are highlighted in blue, with the shade of the highlight representing the magnitude of the contribution. For very small coefficients, the highlight may not be visible.

\paragraph{Example-based explanations} 
As discussed in Section~\ref{related:xai},
example-based explanations may include either representative prototypes of the class that the AI system predicts for the given instance or examples that are similar to the given instance along with their AI predictions and ground truth labels. 
We focus on the latter as they are thought to better help people understand how the model makes mistakes when its prediction differs from the ground truth~\cite{lai2021towards}. While there exist more sophisticated approaches to select examples (e.g., specifically looking for examples with different ground truth labels~\cite{lai2019human,wang2021explanations} or using influence functions~\cite{koh-icml2017}), we opt for a simple method of selecting the two nearest neighbors from the training set. 
In the income prediction task, we display the two individuals from the training set with the closest Euclidean distance to the current individual, with categorical features one-hot encoded, as shown in Figure~\ref{fig:example_explanations} (b). 
In the biography classification task, we display the two nearest neighbors with the smallest Euclidean distance to the given biography using the bag-of-words embedding (i.e.,  the same feature space used by the underlying model), as shown in  Figure~\ref{fig:example_explanations} (d).
If participants used the nearest neighbor's ground truth labels to predict the outcome for the current instance, they would achieve 63\% accuracy on the income prediction task and 60\% accuracy on the biography classification task. This means that to get a performance boost from example-based explanations, participants would need to perform non-trivial reasoning, beyond simply relying on the ground truth of the nearest neighbors.

\subsection{Experimental Design}

We used a within-subjects design to study the effect of explanation type. Each participant was randomly assigned to complete either the income prediction task or the biography classification task, but engaged with both explanation types. This allowed us to directly ask participants to compare the two types of explanations in the post-study interview.  To account for ordering effects, we randomized which explanation type was shown to each participant first.
We ensured that half of the participants were assigned to each task and approximately half of those participants saw each explanation type first.
Note that we do not intend to make direct comparisons between participants' behavior in the two tasks since the tasks differ along many dimensions (e.g., data modality, domain, 
ML model accuracy, user interface). However, including both tasks helps us to understand the generalizability of our observations.

Prior works that involved controlled experiments on human-AI decision-making have typically adopted one of two workflows~\cite{lai2021towards}: The participant either saw a prediction from an AI system and then made their own decision, or the participant made their own decision first and then was given an opportunity to update their decision once they were shown the AI prediction. One advantage of the latter workflow is that it allows for studying how a participant's behavior differs for instances where they initially agree or disagree with the AI prediction. It also allows a baseline measure of participants' performance without AI support.  The drawback is that it may not resemble common real-world use cases of AI decision-support, and asking participants to state their own decision immediately before seeing the AI system's may inhibit their tendency to follow the AI prediction.  To balance these benefits and drawbacks, we first asked participants to complete all instances of the decision task on their own without AI support and then showed participants the same set of instances with AI support (in random order).

Each participant was asked to make predictions on 16 instances selected from a large test bank with stratified sampling---10 from those instances where the model made correct predictions and 6 from those where it made incorrect predictions. This means that while the accuracy of the models on held-out data was higher (80.8\% for income prediction and 75.4\% for biography classification, as described above), participants' ``experienced accuracy''  was only 62.5\%.  We over-sampled instances on which the AI prediction was wrong so that we would be better able to explore whether and how participants would over-rely on the AI system. 
We chose this sampling strategy rather than training a worse-performing model to begin with because the explanations and errors made by a worse-performing model would not be realistic for an acceptable decision-support AI system.

Our above experimental design was informed and refined based on a pilot study with 5 participants. 
The goal of the pilot was to understand (1) how many decisions participants could make within 30 minutes and if they experienced decision fatigue after the allotted time, (2) whether the study workflow felt unnatural to participants (e.g., if they were bothered by the repeated  tasks), and (3) whether the instructions provided about the task and explanations were easy to follow.
Participants in the pilot study (and in the final study) often did not recognize that instances were repeated or could not immediately recall what decisions they made without the AI system, perhaps due to the complexity and quantity of instances they saw,
which we believe mitigates the concern about the unnaturalness of our workflow with repeated tasks.

\subsection{Procedure}

The study was conducted online. Prior to participating, each participant filled out a consent form in which they were asked to consent to their camera and screen being recorded during the study. At the start of the study, participants were asked to enter a video conferencing platform, check their microphone, speaker, and camera, and share their screen with the moderator (the first author). 
The remainder of the study was divided into four phases, as described below. 
The study took an average of 45 minutes to finish and each participant was compensated with \$35.

\paragraph{Phase 1: Practice with the task.} In order to familiarize participants with their decision task (either income prediction or biography classification) and give them a reasonable sense of the task domain, participants were first asked to make decisions on 5 randomly sampled instances of their task. After each decision, they were provided with feedback on whether they were correct, exposing them to ground truth labels for the task.

\paragraph{Phase 2: Human-alone decision-making.} Participants were then asked to make decisions on an additional 16 instances without receiving feedback. As described above, this allowed us to determine which instances the participant initially agreed or disagreed with the AI prediction on. It also let us measure the participant's baseline accuracy on the task with no AI assistance.

\paragraph{Phase 3: AI-supported decision making.} In this phase, participants engaged with the two types of explanations (feature-based and example-based) in random order. For each explanation type, the participant was first introduced to the type of explanation with an illustrated example, as shown in Appendix~\ref{appdx:instructions}, and then asked to make 8 decisions while seeing the AI system's prediction and explanation, as shown in Figure~\ref{fig:example_explanations}.  The 16 instances shown in this phase were the same 16 instances used in phase 2 (order randomized).
As discussed, for each explanation type, we sampled 5 instances on which the AI prediction was correct and 3 instances on which it was incorrect.

During this phase, we asked participants to think aloud by vocalizing their thought process as they looked at each instance, engaged with the provided information, and made their decision. 
The moderator prompted participants who were less vocal to share their thought processes. (We did not require participants to think aloud in earlier phases of the study to save time and avoid fatigue.)

\paragraph{Phase 4: Post-task interview.} We closed the study with a brief interview to understand the participant's experience making decisions with both types of explanations and how they thought the system could be improved. The full set of interview questions is available in Appendix~\ref{appdx:interview}.

\subsection{Analysis Approach}
\label{sec:analysis}

We used a mix of quantitative and qualitative methods to analyze the study data.  On the quantitative side, we performed an exploratory analysis to study the effect of explanation type on participants' accuracy (RQ2). We describe the specific analyses we ran when we discuss the results in Section~\ref{sec:rq2}.

We collected two types of data for qualitative analysis: participants' think-aloud data and their responses from the post-study interviews. We analyzed the think-aloud data in order to answer both RQs. 
The first author and second author first performed open coding informed by Grounded Theory research~\cite{walker2006grounded} on a common set of four participants' data. They iteratively discussed and developed a set of common codes and themes (discussed in Section~\ref{result:common-themes}), and then  each coded the think-aloud data from half of the remaining participants. Specifically, for each decision made by each participant, they coded what happened during the decision process, and recorded the codes in one row in a spreadsheet. In this step, the coders were blind to whether the decision was correct. After all decision tasks were coded in this manner, the authors filled in columns about the correctness of participants' original and final decisions, and the correctness of the AI's predictions. We then conducted a \textit{comparative analysis} by separating cases in which we observed appropriate or inappropriate reliance on the AI system, and contrasted codes for participants using different explanations. This comparative analysis allowed qualitative insights into how these explanations affect reliance differently (RQ2), as discussed in Section~\ref{sec:rq2}.

We also analyzed the interview data to understand participants' subjective perceptions. The first author followed the interview structure and extracted themes around participants' perception of the two types of explanations and how they wish to improve them, as discussed in Section~\ref{qual:improvement}.

\section{Results} 

First, we overview themes from the think-aloud data that reflect common elements in participants' decision-making processes with AI support, focusing on the types of intuition participants brought into the process (RQ1). Next, we quantitatively analyze the effect of explanation type on decision accuracy and reliance and use think-aloud data to explore the reasons why the two types of explanations had different effects (RQ2). Lastly, we discuss participants' subjective perceptions of the explanations and the improvements they suggested during the post-study interviews.

\subsection{Common Types of Intuition Applied in the Decision-Making Process (RQ1)} 

\label{result:common-themes}

We present common themes identified from participants' think-aloud data to answer RQ1: What types of human intuition are involved in engaging with AI predictions and
explanations and how do they affect reliance on AI?
 For each theme, we first give an overview, and then briefly discuss how the two types of explanations differ around the theme (RQ2). 
We then answer the second part of RQ1 by discussing how these themes suggest what we refer to as \emph{intuition-driven pathways} for decision-makers
to override the AI prediction when they believe it is wrong.
We further delve into how these pathways explain the different effects of the two types of explanations in Section~\ref{sec:rq2}.

\paragraph{\textbf{Intuition about the outcome}}  Participants commented on their own intuition or ``gut feeling'' about what the decision outcome should be---whether the person more likely made above or below \$50,000 in income prediction task, and which profession the person more likely had in biography classification task---sometimes \emph{before even considering the AI prediction and explanation}. While our experiment setup did not explicitly isolate participants' own intuition (in phase 3) from the impact of AI output, we observed that some individuals intentionally chose to make their own judgment first before attending to AI outputs.\footnote{All participants of course made their own judgments first during phase 2, but as mentioned above, we did not collect think-aloud data during this phase in order to avoid participant fatigue.}
Some participants explicitly commented on their ``\textit{first reaction},'' such as ``\textit{it is medicine [related]...it doesn't really sound like a professor [as AI predicts], so I'll read down through [the examples]}'' (P31).

Participants' comments also reflected the \emph{strength} of their intuition about the outcome (or confidence~\cite{lu2021human}), which impacted how they engaged with the AI system. When their intuition about the outcome was strong, participants discounted the AI system's output. For example, seeing an individual with the occupation of a truck driver, P11 said ``\textit{I actually feel like this is a case where I have a little background and that really contradicts with the AI [prediction]. If truck drivers are doing private goods hauling, they are actually paid quite a lot because they have a pretty tough job... So I'm actually going to ignore the AI prediction.}''  
When their strong intuition agreed with the AI system's prediction, participants more readily and quickly made a decision, sometimes even without checking the explanations. 

In contrast, participants who acknowledged their intuition about the outcome was weak tended to examine the explanations more closely, hoping to find additional information to help them judge the correctness of AI prediction. If they found no evidence that the AI prediction might be wrong or they simply could not reason about the explanation meaningfully, they tended to defer to the AI prediction. For example, P23 spent a long time looking at the feature-based explanations and said ``\textit{Doesn't seem unbelievable to me... I'm a little less familiar with this occupation... and I have a lot more uncertainty around what would be a good [indicator] here. And so I think I'm more comfortable relying on the AI prediction.}''

As illustrated in P11's quote above, participants may form their intuition about the outcome by retrieving prototypes or similar examples from previous experiences, which were often prompted by one or a subset of features that caught their attention, such as a profession they know about. 
In some cases, participants anchored their judgment on exceptional or rare feature values. When P28 noticed a given individual worked 55 hours per week, they immediately said ``\textit{I'll go with above [\$50,000] just based on the hours worked per week.}'' Occasionally, participants were influenced by a similar example that they saw earlier in the practice phase.

\textit{How did intuition about the outcome differ across the two types of explanations?} 
Interestingly, we observed that participants were \textit{more likely to acknowledge that they had weak intuition about the outcome with feature-based explanations}.
One reason may be that feature-based explanations are more prominently displayed, 
particularly in the biography classification task where the explanation is visually overlaid on top of the text as highlights, which is typical for explaining text data. For example, P31 complained that the text highlights made them ``\textit{think less}'' even when ``\textit{non-informative}'' words were highlighted. P22 said ``\textit{my approach with it in this is to scan the pink, scan the blue [instead of] read more. That's what this interface is leading me to do.}''  
In contrast, example-based explanations allowed participants to easily focus on the current instance first before attending to the explanation, and even to ignore the explanation altogether when they had a strong intuition about the outcome.

\paragraph{\textbf{Applying intuition about features to reason about explanations}} 
While participants paid attention to prominent features to arrive at their own intuition about the outcome, as described above, a significant portion of the think-aloud data included participants' comments on their intuition about features when \textit{engaging with explanations}. This suggests that showing explanations created an additional step in the decision-making process, i.e., to judge the impact of features that had not been previously considered or to reason more precisely about certain features. For example, even though P9 came to the same prediction as the AI system, seeing the explanation prompted an additional comment ``\textit{[the explanation] is really picking up on age a lot. I agree that it is important, but I'm surprised by the magnitude.}''
When reasoning about explanations, participants would look for evidence that indicated an incorrect model prediction, but sometimes ended up updating their own intuition about features when doing so.

We first summarize the types of intuition that participants applied when reasoning about explanations, then elaborate on how they were used differently for the two types of explanations. 

\begin{itemize}
    \item \textit{Feature relevance and weights}. Participants most frequently commented on the weight of a feature presented in the explanation such as ``\textit{it is saying occupation is a big deal, and I agree...engineering would be a positive}'' (P32). With text data, they tended to talk about binary ``relevance.''
     Explanations also prompted some participants to comment on the relative weights of multiple features. For example, P1 and P23 believed age should have a higher weight than education in determining income level.

    \item \textit{Interaction between features}. Participants also reacted to features in relation to each other, bringing in nuanced domain knowledge, such as ``\textit{self-employed, not sure if that would necessarily make the most difference because this is a trade occupation}'' (P23), or ``\textit{adjunct clinical lecturer---the model did not pick [adjunct] up but `university.' That indicates professor is not the actual profession.}'' (P31).

    \item \textit{Infer additional features}. For income prediction, we observed that participants inferred additional features by combining existing features,  such as inferring an individual's ``\textit{career stage}'' (P26) based on the combination of education, age, and profession, then using that inference to challenge the AI explanation that under-weighted age. This pattern was even more frequent in biography classification, where participants not only utilized combinations of words and phrases, but also higher-level features such as ``\textit{this type of art journals},'' or ``\textit{academic associations}'' (P30), as well as meta-features like the writing style and format.

    \item \textit{Assign granular or different meanings to a feature and judge weights accordingly}. Interestingly, we observe some participants assigned a granular or different meaning to a feature based on their pre-existing beliefs or prototypical cases they could recall. For example, when seeing the occupation ``engineers,'' multiple participants attempted to assign a specific type of engineer to reason about the feature weight, as ``\textit{there [are] maybe trigger words in my brain that make me to believe that it's a certain job}'' (P32).
    \item \textit{Assign granular meaning to a label and update feature weights accordingly}. 
    For example, after seeing the AI system predict professor and also highlight medical-related keywords, P31 realized that ``\textit{it could be medical professor}'' and accepted the AI explanation and prediction. 
\end{itemize}

\textit{How did intuition about features differ across the two types of explanations?} With feature-based explanations, participants applied intuition about features to determine if they agreed or disagreed with the model's reasoning. \textit{Feature information they disagreed with is considered evidence that discredits the model's prediction}.

When given example-based explanations, intuition about features was most commonly used in two ways:
First, intuition about features, including all the types discussed above, guided evaluations of whether a given instance was indeed similar to the provided examples.  
For example, P4 judged an example to be similar based on features that they believe should carry more weight: ``\textit{The [given individual] works a similar number of hours to [similar] profile 1 and has a similar education background.}'' P32 judged an example to be dissimilar based on the interaction of features: ``\textit{at their age, an associates degree versus bachelors degree can matter}.'' 
Second, intuition about features affected reasoning about the impact of feature values that differed between the instance and the example to infer the likely outcome. For example, P1 looked at the examples and said ``\textit{similar profile and also this person is hard work with more hours. So I chose [to go with more than \$50,000].}'' 
Both judgments can help identify evidence to discredit the AI prediction, \textit{either by confirming that the instance is similar (dissimilar) to an example that the model predicted incorrectly (correctly) or by identifying features with different values that indicate a different outcome than the prediction}. 
Additionally, participants were \textit{more likely to identify and infer the impact of new features} with example-based explanations through reasoning about the similarity of examples. For example, P22 noticed an organization shared by one example and learned that this is ``\textit{some teaching thing...book club}'' and then switched from their initial intuition of ``psychologist'' to ``teacher.''
This suggests that example-based explanations can provide additional context and support inductive reasoning~\cite{buccinca2020proxy} to help people form additional intuition about features.

\paragraph{\textbf{Intuition about AI limitations and prediction unreliability}} 
Lastly, we observed comments on the limitations of the AI system, particularly when participants recognized signals revealed by the explanations that indicated the \emph{unreliability} of the model's prediction on the current instance. While previous work suggests that a desideratum of XAI methods should be uncertainty awareness~\cite{wang2021explanations,carvalho2019machine}, we avoid using the term ``uncertainty'' as participants' comments reflect their subjective perception of reliability rather than a measure of the actual model uncertainty.  Recognizing that a prediction was unreliable could boost a participant's outcome intuition if they disagreed with the AI prediction, or prompt self-doubt and further deliberation if they initially agreed.

Participants, primarily those with ML experience, also mentioned other general limitations of AI to justify cases where they discounted the AI prediction. One common intuition was regarding biases that can be embedded in ML models, which led participants to discount predictions in cases where the feature-based explanation gave a high weight to gender or in cases where the example-based explanation showed different predictions for examples with different genders. Another common intuition is AI's limitation in considering contexts or multiple features. For example, P9 commented that ``\textit{once it's like in the context of this occupation, the contribution [of other features should] change. I don't know if that happened in this case for the model.}'' 
Lastly, participants commented that AI might not be good at predicting rare instances, such as biographies with an uncommon format.

\textit{How did intuition about AI limitations differ across the two types of explanations?} We found participants utilized different signals of prediction unreliability with different frequencies for the two types of explanations. With the example-based explanation, most participants identified prediction unreliability in instances where the AI prediction was incorrect on similar examples,
such as when ``\textit{the [AI] predictions [of the examples] are completely off from the actual income}'' (P32). 
With the feature-based explanation, (only) a few participants noted a pattern of unreliability when the weights did not trend strongly in either direction. For example, in the tabular setting, participants noted when there were fairly equal numbers of features providing evidence in both directions, or when the scores were ``\textit{pretty uniform}'' [P17].

\paragraph{\textbf{Summary: Three Intuition-Driven Pathways to Non-reliance on AI}} In light of prior findings showing explanations lead to overreliance on AI predictions, we summarize three pathways that participants used in our study to apply different types of intuition to override the AI prediction.
\begin{itemize}
    \item \textbf{Pathway 1:} Form a strong intuition about the outcome that disagrees with the AI prediction.
    \item \textbf{Pathway 2:} Apply intuition about features to reason about the AI explanations and identify evidence that discredits the AI prediction.
    \item \textbf{Pathway 3:} Recognize AI limitations, especially signals of prediction unreliability.
\end{itemize}{}

We note that these pathways are identified from reasoning that was frequently mentioned in participants' think-aloud data, which does not allow isolating or quantifying their effect. Therefore, these pathways should not be taken as mutually exclusive or having different levels of impact. Reasoning along these pathways also does not imply that non-reliance is appropriate.

\subsection{The Effect of Explanation Type on Accuracy and Reliance (RQ2)}\label{sec:rq2}

We now evaluate whether the choice of feature- or example-based explanations leads to differences in decision accuracy and appropriate reliance, and if so, whether this can be explained by differences in how intuition comes into play (RQ2).
To do so, we first quantitatively analyze the decision outcomes from our study. We then explain the observed effects using insights from qualitative analysis, through the lens of the three intuition-driven pathways identified above.
Given the small sample size and that the focus of our study is not on hypothesis testing, we consider the quantitative analysis exploratory. We encourage readers to use caution when interpreting the $p$-values and focus instead on the trends.  

\subsubsection{Exploratory quantitative analysis} \label{sec:quant_analysis}

\begin{figure}[th!]
     \centering
     \begin{subfigure}[b]{0.48\textwidth}
         \centering
         \includegraphics[width=0.99\textwidth]{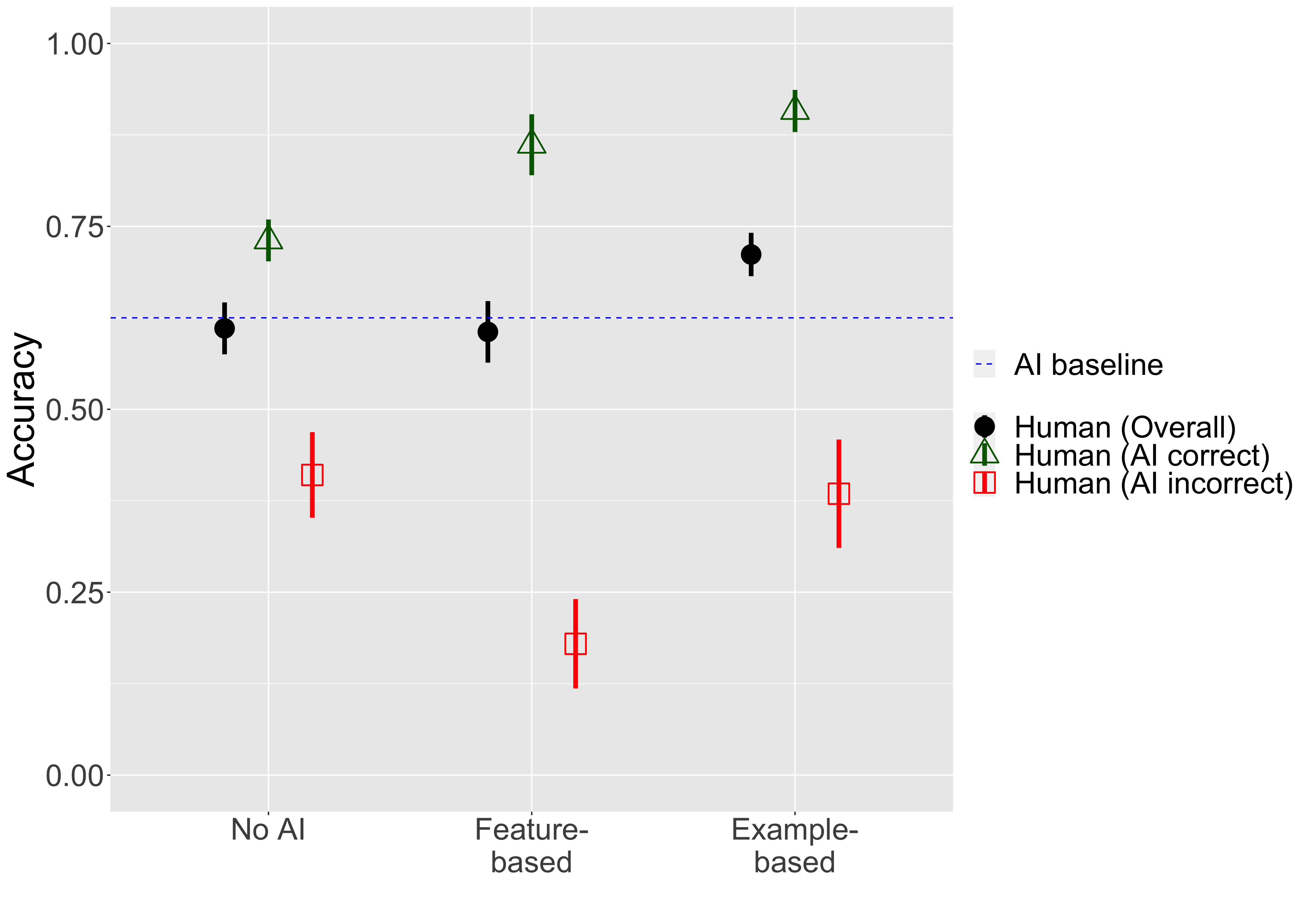}
         \caption{Income Prediction}
         \label{fig:agg_perf}
     \end{subfigure}
     \hfill
     \begin{subfigure}[b]{0.48\textwidth}
         \centering
         \includegraphics[width=0.99\textwidth]{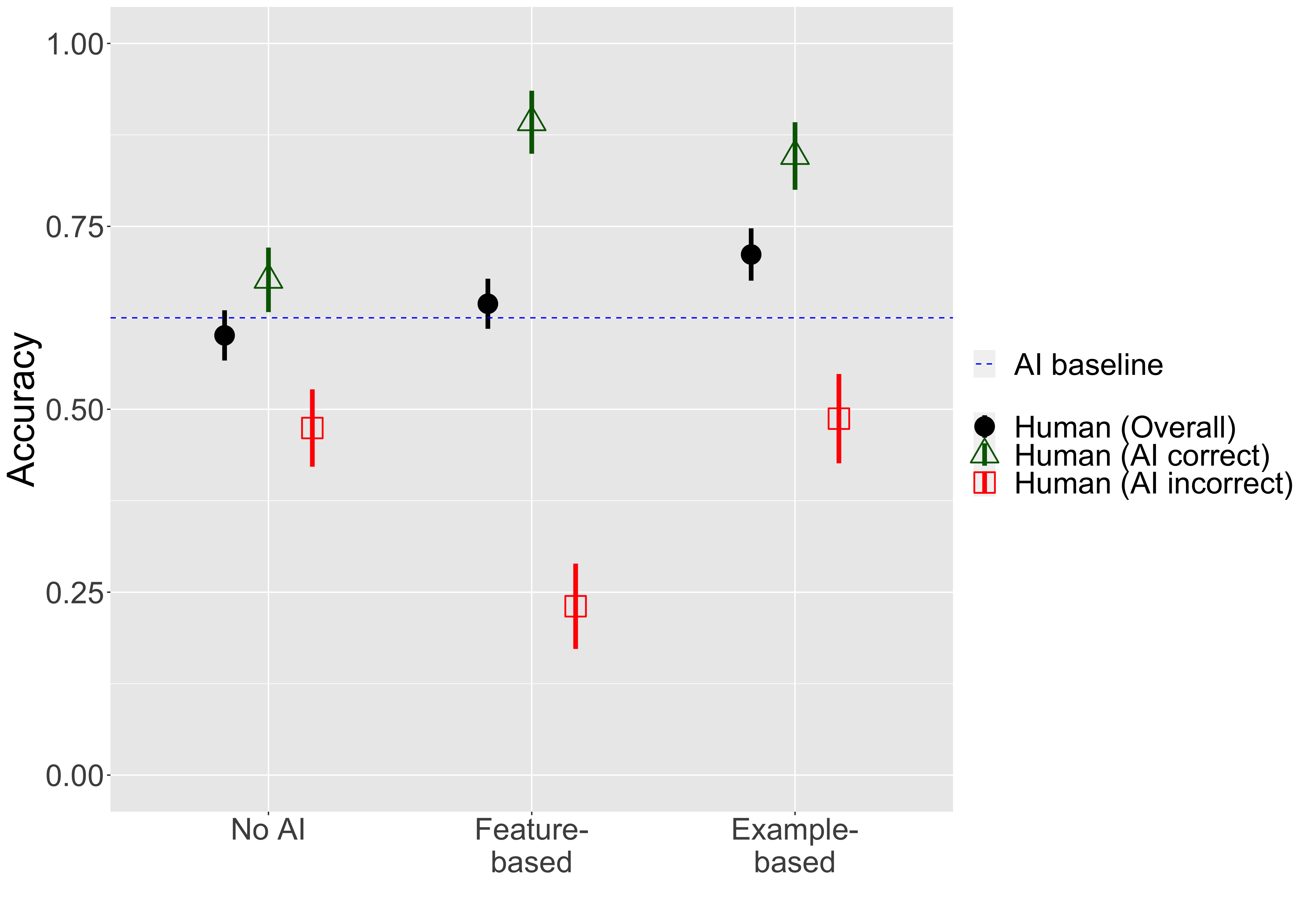}
         \caption{Biography Classification}
         \label{fig:AIcorrect_perf}
     \end{subfigure}
     \caption{Participant's accuracy in different conditions---no AI (from phase 2), AI support with example-based explanations (from phase 3), and AI support with feature-based explanations (from phase 3)---compared to the accuracy of the AI baseline on the two prediction tasks.
     Across tasks, example-based explanations achieved complementary decision performance whereas feature-based explanations did not.
     \label{fig:accuracy}}
\end{figure}

To begin, we compare participants' decision accuracy across explanation types, where accuracy is the percentage of instances for which a participant's decision matches the ground truth label.  The decisions we look at are the participant's final decisions in phase 3 of the study (8 instances for each explanation type).  As a baseline, we also compute the accuracy of participants' decisions in phase 2 (16 instances), which we refer to as the ``No AI'' condition.  Figure~\ref{fig:accuracy} shows the mean and standard deviation of accuracy for each of these conditions, as well as the accuracy for cases in which the AI prediction was correct or incorrect, respectively, as we discuss more below.

The average overall accuracy and standard error of participants with no AI, AI support with feature-based explanations, and AI support with example-based explanations were $61.1\% \pm 2.9\%$, $60.6\% \pm 4.2\%$, and $71.1\% \pm 2.9\%$ respectively in the income prediction task,  and $60.0\% \pm 4.4\%$, $64.4\% \pm 4.3\%$, and $71.2\% \pm 4.6\%$ 
respectively in the biography classification task  (Figure~\ref{fig:accuracy}, in black). Since the accuracy of the models we trained on the particular instances presented in the study is 62.5\%, this means that the example-based explanations led to \textit{complementary performance}---accuracy higher than human or AI alone---on both tasks.

We ran a mixed-effect regression on the decision accuracy, using participants as a random-effects variable, and explanation type as a fixed-effects variable,
where No AI is used as the reference level. We find that feature-based explanations did not have a significant effect on decision accuracy over No AI for either task ($SE=0.00, p=0.88$ for income prediction and $SE=0.04, p=0.51$ for biography classification), whereas the effect of example-based explanations is marginally significant ($SE=0.10, p=0.07$) for income prediction  and significant ($SE=0.11, p=0.04$) for biography classification.

To better understand the effect of explanation type on \textit{appropriate reliance}, we separately analyze accuracy for instances on which the AI system was correct and instances on which it was incorrect. For cases where the AI system made correct predictions, consistent with prior work \cite{wang2021explanations}, we find that both types of explanations led to increased accuracy on both tasks (Figure~\ref{fig:accuracy}, in green).  Running an analogous regression analysis to the one for overall accuracy above, we find that, for these AI-correct cases, feature-based explanations had a marginally significant effect on accuracy ($SE=0.13, p=0.06$) for income prediction and a significant effect on accuracy  ($SE=0.22, p=0.01$) for biography classification. Example-based explanations had a significant effect ($SE=0.18, p=0.01$) for income prediction and a marginally significant effect ($SE=0.17, p=0.06$) for biography classification.

For cases where the AI system was incorrect, consistent with prior work finding that feature-based explanations could reduce decision accuracy by increasing people's overreliance~\cite{zhang2020effect,bansal2021does,poursabzi2021manipulating}, we find that feature-based explanations led to decreased accuracy compared with the No AI condition. This effect is marginally significant ($SE=-0.23, p=0.09$) for income prediction and significant ($SE=-0.24, p<0.01$) for biography classification. In contrast, we find no such effects for example-based explanation ($SE=-0.03, p=0.80$ for income prediction and $SE=0.01, p=0.43$ for biography classification); participants were able to maintain a similar accuracy as they had without AI support in these cases. These results are illustrated in Figure~\ref{fig:accuracy} in red.

In summary, our analyses echo prior work and suggest that \textbf{feature-based explanations may not improve decision accuracy because they increase participants' reliance on AI} regardless of whether the AI system is correct or incorrect. In contrast, \textbf{example-based explanations helped achieve complementary human-AI performance} in our study by increasing appropriate reliance when the AI system was correct while helping participants maintain their accuracy when the AI system was incorrect. Next, we further delve into the reasons for these observations.

\subsubsection{Why did example-based explanations better support decision-making than feature-based explanations?} \label{result:reliance}
To answer this question, we investigate participants' reliance on the AI system through the lens of the three intuition-driven pathways identified in Section~\ref{result:common-themes}. 
To do this, we further break down participants' decisions into four different scenarios based on whether the participant's initial decision in phase 2 was correct (denoted P2 \cmark) or incorrect (P2 \xmark), and whether the AI prediction was correct (denoted AI \cmark) or incorrect (AI \xmark).
Table~\ref{tab:switching} shows participants' phase 3 accuracy in each of these scenarios.

\begin{figure}[t!]
     \centering
     \begin{subfigure}[b]{0.48\textwidth}
         \centering
         \includegraphics[width=0.99\textwidth]{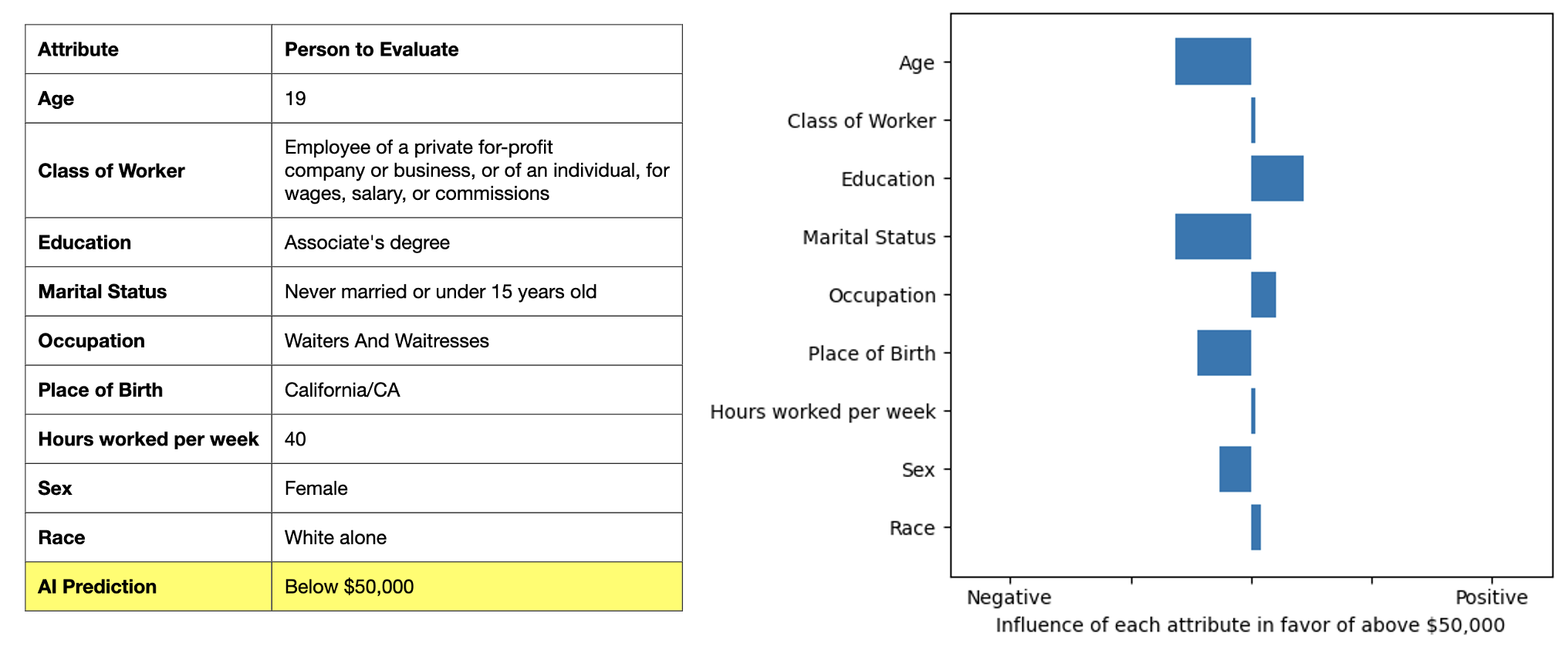}
         \caption{Example of a \emph{Low Diff} feature-based explanation for income prediction.}
         \label{fig:tab_unreliable1}
     \end{subfigure}
     \hfill
     \begin{subfigure}[b]{0.48\textwidth}
         \centering
         \includegraphics[width=0.99\textwidth]{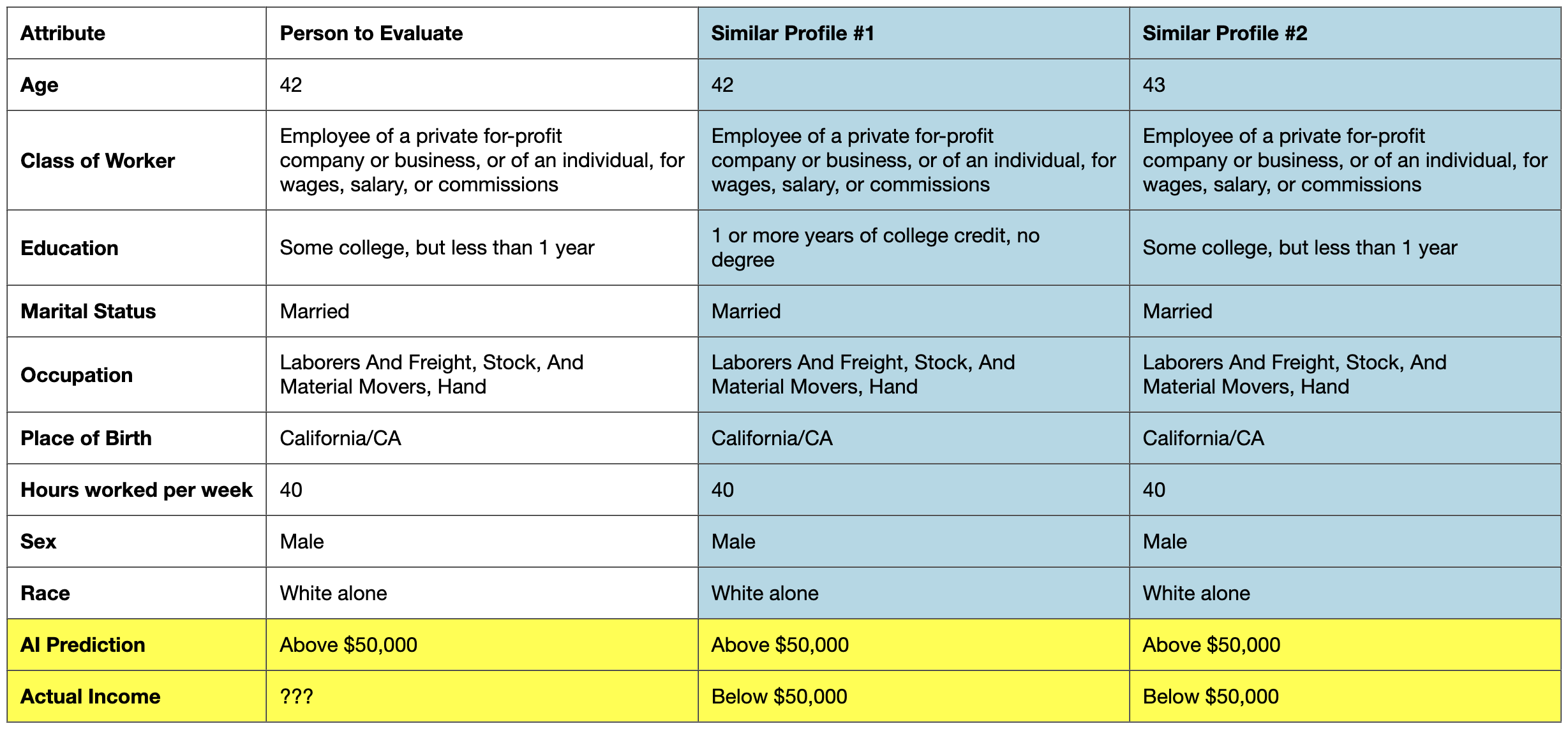}
         \caption{Example of a \emph{Both Wrong} example-based explanation for income prediction.}
         \label{fig:tab_unreliable2}
     \end{subfigure}
          \centering
     \begin{subfigure}[b]{0.48\textwidth}
         \centering
         \includegraphics[width=0.99\textwidth]{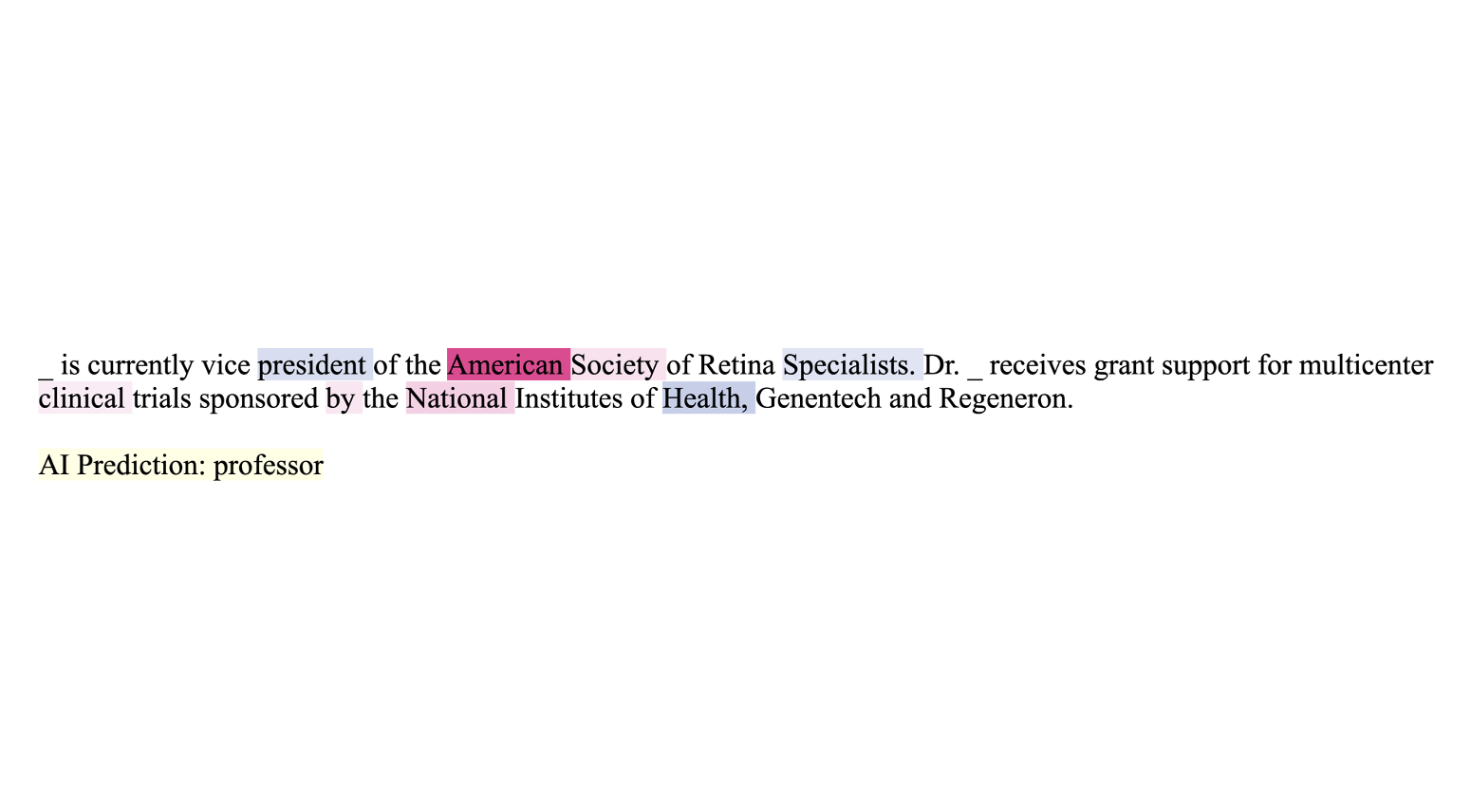}
         \caption{Example of a \emph{Low Diff} feature-based explanation for biography classification.}
         \label{fig:text_unreliable1}
     \end{subfigure}
     \hfill
     \begin{subfigure}[b]{0.48\textwidth}
         \centering
         \includegraphics[width=0.99\textwidth]{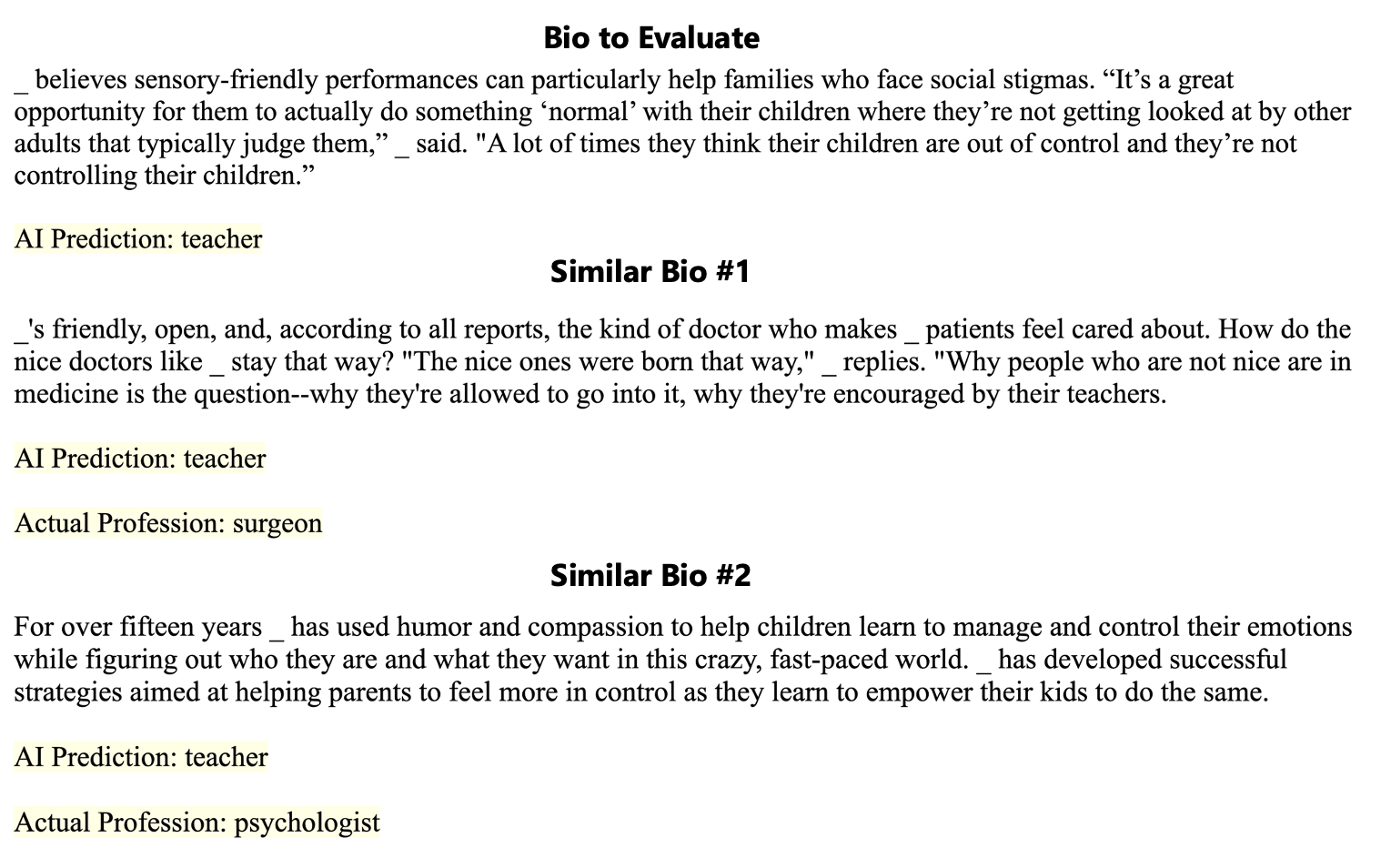}
         \caption{Example of a \emph{Both Wrong} example-based explanation for biography classification.}
         \label{fig:text_unreliable2}
     \end{subfigure}
     \caption{Examples of feature- and example-based explanations that signal prediction unreliability.}
     \label{fig:unreliability}
\end{figure}

\paragraph{Quantifying the existence of unreliability signals in explanations} To explore the effect of pathway 3---how explanations help participants recognize prediction unreliability---we define a way of measuring whether a particular explanation signals unreliability.
For feature-based explanations, as discussed in Section~\ref{result:common-themes}, participants mentioned using the fact that the total feature weights did not trend strongly in either direction as a signal of an unreliable prediction. 
To capture this, we heuristically define the following proxy measure, which we refer to as \emph{Low Diff}. First, we calculate the absolute difference between the sum of positive weights and the sum of negative weights for each instance used in the study.  We use the value of the 25th percentile of the absolute differences 
as a threshold, and consider any instance where this difference is lower than the threshold as  signaling unreliability. We refer to instances where the difference is above the threshold as \emph{High Diff}.  Examples of \emph{Low Diff} explanations for the two tasks are shown in Figure~\ref{fig:unreliability} (a) and (c), while the explanations in Figure~\ref{fig:example_explanations} (a) and (c) are \emph{High Diff}.

\begin{table}[htb]
    \caption{For each task (income prediction, top and biography classification, bottom) and each type of explanation (feature-based and example-based), we break down participants' accuracy by four scenarios that capture different forms of (non-)reliance based on whether their initial decision in phase 2 was correct (denoted by a \cmark in the P2 column) or incorrect (\xmark) and whether the AI prediction was correct (denoted by a \cmark in the AI column) or incorrect (\xmark). For each scenario and each phase 3 decision (P3 column) we additionally report the number of times that the provided explanations did or did not exhibit signals of unreliability.}
{\begin{tabular}{cc|ccc|cccc}
\hline
\multicolumn{9}{c}{\textbf{Income Prediction Task}} \\
& & \multicolumn{3}{|c}{\textbf{Feature-Based}} & \multicolumn{4}{|c}{\textbf{Example-Based}}  \\
(AI, P2) & P3 & Accuracy   & \multicolumn{1}{c}{\begin{tabular}[c]{@{}c@{}}High\\ Diff\end{tabular}} & \begin{tabular}[c]{@{}c@{}}Low\\ Diff\end{tabular}  & Accuracy  & \begin{tabular}[c]{@{}c@{}}Both\\ Right\end{tabular} & Mixed & \begin{tabular}[c]{@{}c@{}}Both\\ Wrong\end{tabular} \\ \hline
 \multirow{2}{*}{(\xmark, \cmark)} & \ \cmark  & \multirow{2}{*}{42.9\%}  & 3 & 3  & \multirow{2}{*}{50.0\%}    & 2     & 1 & 6   \\ 
& \xmark  &     & 6 & 2     &   & 4 & 4 & 1    \\ \hline
 \multirow{2}{*}{(\xmark, \xmark)} & \cmark & \multirow{2}{*}{4.0\%}  & 0 & 1 & \multirow{2}{*}{28.6\%}  & 1 & 0 & 5  \\
 & \xmark  &     & 10 & 14     &   & 10 & 4 & 1   \\ \hline
 \multirow{2}{*}{(\cmark, \cmark)} & \cmark & \multirow{2}{*}{91.7\%}  & 32 & 12 & \multirow{2}{*}{100.0\%}  & 37 & 10 & 0  \\
 & \xmark  &     & 4 & 0     &   & 0 & 0 & 0    \\ \hline
 \multirow{2}{*}{(\cmark, \xmark)} & \cmark  & \multirow{2}{*}{70.6\%}  & 11 & 1& \multirow{2}{*}{66.7\%}  & 12 & 0 & 0 \\
 & \xmark   &     & 5 & 0    &   & 0 & 6 & 0    \\ 
\ \\
\ \\ 
\hline
\multicolumn{9}{c}{\textbf{Biography Classification Task}} \\
& & \multicolumn{3}{|c}{\textbf{Feature-Based}} & \multicolumn{4}{|c}{\textbf{Example-Based}}  \\
(AI, P2) & P3 & Accuracy   & \multicolumn{1}{c}{\begin{tabular}[c]{@{}c@{}}High\\ Diff\end{tabular}} & \begin{tabular}[c]{@{}c@{}}Low\\ Diff\end{tabular}  & Accuracy  & \begin{tabular}[c]{@{}c@{}}Both\\ Right\end{tabular} & Mixed & \begin{tabular}[c]{@{}c@{}}Both\\ Wrong\end{tabular} \\ \hline
 \multirow{2}{*}{(\xmark, \cmark)} & \ \cmark  & \multirow{2}{*}{41.2\%} & 4 & 3  & \multirow{2}{*}{75.0\%}    & 5   & 3 & 7   \\ 
& \xmark  &     & 4 & 6     &   & 2 & 2 & 1    \\ \hline
 \multirow{2}{*}{(\xmark, \xmark)} & \cmark & \multirow{2}{*}{9.1\%}  & 0 & 2 & \multirow{2}{*}{21.1\%}  & 1 & 1 & 2  \\
 & \xmark  &     & 10 & 10     &   & 8 & 7 & 0   \\ \hline
 \multirow{2}{*}{(\cmark, \cmark)} & \cmark & \multirow{2}{*}{95.7\%}  & 42 & 3 & \multirow{2}{*}{100.0\%}  & 29 & 12 & 0  \\
 & \xmark  &     & 2 & 0     &   & 0 & 0 & 0    \\ \hline
\multirow{2}{*}{(\cmark, \xmark)} & \cmark  & \multirow{2}{*}{72.2\%}  & 8 & 5& \multirow{2}{*}{58.3\%}  & 9 & 5 & 0 \\
 & \xmark   &     & 3 & 2    &   & 6 & 4 & 0    \\ \hline
\end{tabular}}
\label{tab:switching}
\end{table}

For example-based explanations, participants tended to notice whether the AI system made mistakes on the two examples provided. We define two heuristic measures of whether an explanation signals unreliability based on whether the system made incorrect predictions on both (\emph{Both Wrong}) or one (\emph{Mixed}). We refer to instances where the AI system made only correct predictions as \emph{Both Right}.  Examples of \emph{Both Wrong} explanations are shown in Figure~\ref{fig:unreliability} (b) and (d), while Figure~\ref{fig:example_explanations} (b) and (d) are examples of \emph{Both Right}.

Note that these measures only reflect the theoretical presence of unreliability signals and do not imply that all participants recognized these signals. As discussed, there were significantly fewer participants who commented on such signals for feature-based explanations.

\paragraph{What happened in these four scenarios? } We now discuss observations about each of the four scenarios, referencing the results summarized in Table~\ref{tab:switching} combined with observations from the comparative analysis of the think-aloud data, as described in Section~\ref{sec:analysis}. For each scenario, we further break down cases by whether the participants' final decision was correct (denoted by a \cmark\ in the P3 column) or incorrect (\xmark\ for P3). For each scenario, we include columns that show the number of cases in which the explanation exhibited signals of (un-)reliability as described above.

\paragraph{AI \xmark, P2 \cmark} Participants' accuracy is higher in this scenario (row 1 in Table~\ref{tab:switching}) than it is when both the AI prediction and their own decisions were wrong (row 2) across tasks and explanation types. This supports the idea that \textit{participants' intuition about the outcome (pathway 1) played a role in their correct non-reliance on AI}. However, in both tasks, participants had higher accuracy with example-based explanations compared with feature-based explanations. 
Across both tasks, for cases where participants correctly overrode the AI prediction (P3 \cmark, first sub-row), there is a high percentage of instances where the AI prediction of the two similar examples were \textit{Both Wrong}.
This suggests that two wrongly predicted examples form a \textit{strong unreliability signal (pathway 3) 
 that played a role in participants' correct non-reliance}. However, having mixed examples did not appear to be a strong enough signal to push participants to not rely on the AI prediction. 

While these statistics only suggest correlations, the think-aloud data support that most participants in this scenario were eager to discredit the AI prediction after noticing it was wrong on both examples. In almost all cases, participants felt more confident about their own intuition and presented their own reasoning to explain the AI system's mistakes. When only one example was wrong, participants more carefully reasoned about the similarity of examples and the impact of different feature values. They had correct non-reliance when they found the instance to be similar to the example, which had different ground truth (pathway 2). Interestingly, in 4 instances of biography classification where participants had correct non-reliance, they commented about \emph{new features} they learned from the examples to help them judge the similarity. For example, P14 correctly disagreed with the AI prediction (teacher) because they noticed that the current profile resembled the two examples (both professors) by all mentioning ``\textit{sophisticated}'' art venues. 

In contrast, for feature-based explanations, there is not a strong correlation between the \emph{Low Diff} proxy measure and participants'
decision correctness. Moreover, there were only two instances where participants noticed this signal of unreliability. When they had correct reliance, they primarily relied on their own intuition (pathway 1) or picked up on feature weights they disagreed with from the explanations (pathway 2). Strikingly, in 5 out of the 8 cases where participants incorrectly relied on the AI prediction for income prediction, they acknowledged they had weak intuition about the outcome and deferred to AI predictions. The remaining 3 went with the AI prediction because they found the explanation agreeable, but only focused on a subset of features. Out of the 10 such cases for biography classification, 3 acknowledged weak intuition, 3 agreed with the explanation, and 4 disagreed with the explanation (e.g., finding highlighted keywords meaningless) but still went with the AI prediction. All these observations support the claim that \textit{feature-based explanations disrupt people's natural intuition} (despite making correct decisions in phase 2), especially when participants think the AI explanation makes sense or are unable to reason about the explanation. 

\paragraph{AI \xmark, P2 \xmark} Interestingly, in this scenario (row 2), while there were very few cases where participants arrived at the right decision with feature-based explanations, example-based explanations surprisingly supported correct decisions for a sizable number of cases. In the majority of these cases, the example-based explanations were \textit{Both Wrong} or \textit{Mixed} (pathway 3). In these cases, we observed that all participants carefully examined the examples to confirm the similarity between the instance and examples with different ground truths (pathway 2), which prompted them to not only override the AI prediction but also their own intuition. In two cases, participants also learned about new features from the examples (pathway 2). In the only three instances where participants were correct with feature-based explanations, they made the correct decision because they disagreed with the feature weights (pathway 2).

\paragraph{AI \cmark, P2 \cmark} Surprisingly, a few participants made incorrect decisions in this scenario (row 3) when using feature-based explanations. This was mainly because they picked up on information they disagreed with in the explanation, e.g., the precise weight of a feature (pathway 2, but incorrect non-reliance). In one case, the participant picked up on the unreliability signal, commenting that there were factors pushing in both directions (pathway 3, but incorrect non-reliance). Even though the AI prediction was correct, there were still a sizable number of cases for which feature-based explanations exhibited \emph{Low Diff}; this is not surprising since instances near the decision boundary can still be predicted correctly.
Meanwhile, example-based explanations had \textit{more informative unreliability signals}---zero cases had both examples predicted wrong. As a result, participants did not have any pathway for non-reliance and reached 100\% accuracy for both tasks. 

\paragraph{AI \cmark, P2 \xmark}
Most participants mentioned their intuition disagreed with the AI prediction  (pathway 1, but incorrect non-reliance) in this scenario (row 4), contributing to lower accuracy compared to row 3. 
This is the only scenario in which feature-based explanations led to higher accuracy than example-based explanations. In all correct reliance cases, participants either found nothing they disagreed with in the feature-based explanation (no pathway 2) or acknowledged they had weak intuition about the decision outcome (no pathway 1). For example-based explanations, none of the cases where participants had correct reliance were cases of \textit{Both Wrong} (no pathway 3), boosting their confidence in the AI prediction.  Considering cases with \emph{Mixed} examples, in the biography classification task, participants changed their decision about half the time, either because they found the example with consistent ground truth to be more similar or they learned new features from the examples. (There were no \textit{Mixed} cases where participants changed decisions in the income prediction task.) 

\ \\
\noindent{These observations about each pathway allow us to summarize why example-based explanations better supported appropriate reliance:} 
\begin{itemize}
    \item For pathway 1 (intuition about the outcome): Example-based explanations led to less disruption of people's natural intuition about the outcome. One reason for this is that they are less visually overwhelming and allow people to focus on the instance first. 
    Additionally, they do not force people to attend to features they would otherwise not pay attention to or quantify the impact, as feature-based explanations often do, which may disrupt or prompt self-doubt about one's own intuition. 
    \item For pathway 2 (intuition about features to reason about explanations): Example-based explanations better supported recognizing or even learning new features because the similar examples and their ground truth labels brought in additional context that could promote inductive reasoning. Reasoning about example-based explanations was also more aligned with people's natural intuition, since it does not require precise quantification of the impact of features. In contrast, when faced with quantified feature importance, participants either found it challenging to judge or difficult to look past information they disagree with, even if it is a relatively trivial disagreement---e.g., ``\emph{the weight for occupation is not high enough}'' (P26). This opened the door to making a wrong decision, even when both the AI prediction and their initial intuition about the outcome were correct.
    \item For pathway 3 (intuition about AI limitations): Example-based explanations provided strong signals of unreliability, particularly when the AI system predicts both examples incorrectly. These signals were not only accurate (highly correlated with AI incorrectness) but were also easily noticeable. These signals helped boost people's outcome intuition when they disagree with the AI prediction and dampen it when they agree with the AI prediction, increasing the chance that people make correct decisions even when both the AI prediction and their own initial intuition are wrong. In contrast, pathway 3 was generally weak for feature-based explanations, at least considering the \emph{Low Diff} proxy measure. 
    \item Differences between the two tasks (tabular vs. text data):  When provided with example-based explanations, participants were more likely to learn about new features from the examples in the biography classification task than in the income prediction task, which we conjecture is due to the biography classification task using text data. This made them more likely to override incorrect intuition even when they were shown examples with mixed ground truth labels (row 4).
    Also, when given feature-based explanations, pathway 2 may have had a slightly weaker impact on the biography classification task, where there were more cases where participants disagreed with the explanation (e.g., picking up on meaningless keywords in the text) but still chose to rely on the AI system.  
\end{itemize}{}

\subsection{Post-Study Interview Responses} \label{qual:improvement}

We analyzed responses to post-study interview questions to understand, subjectively, how participants perceived the two types of explanations and their preferences, and also their suggestions on how to improve both types of explanations in future systems.

\subsubsection{Subjective preferences of explanation types.}
While the empirical findings above show that example-based explanations are better at supporting decision-making, subjectively, participants showed rather mixed preferences: about half favored feature-based explanations (N=11) and the other half preferred example-based explanations (N=10). The remaining participants (N=5) were impartial, noting the potential benefits and drawbacks of each explanation type.

Participants' subjective responses favoring example-based explanations further supported our conclusions in Section~\ref{sec:rq2}. Multiple participants commented that the \textit{unreliability signals or the lack thereof} (i.e., correct predictions on both similar examples) \textit{played a key role} in their confidence in the AI system. For example, P17 said ``\textit{when the data was super consistent, I was gonna go with it}'' (P32). P1 noted that ``\textit{I trusted [ground-truth of the similar examples] more than the predictions of the AI itself}.'' Others liked that the similar examples provide them with \textit{more context to form intuition about features}, such as to ``\textit{compare and contrast [the current input] with the neighbors. For example, I can narrow down what features should I look at}'' (P27), or ``\textit{in breaking ties}'' (P30).

Feature-based explanations were often preferred because subjectively they were \textit{easier to consume}. Some participants found feature-based explanations more intuitive and useful to understand how the AI system made the prediction: ``\textit{seeing if the factors that have contributed most to the decision were aligned with what my thought process was}'' (P9) and the highlights helped form ``\textit{a picture in my mind for how the AI model might have been acting}'' (P31). 
 Interestingly, while some participants found the overlay presentation of feature-based explanations distracting, others found it appealing to ``\textit{guide my attention}'' (P18) or ``\textit{help speed up my scanning process}'' (P8) without realizing the risk of feature-based explanation disrupting them from forming their own intuition. 

We find a disparity between what participants \textit{thought} was helpful and what was actually helpful for their decisions: 7 out of 11 participants who preferred feature-based explanations performed better with example-based explanations.
A recent work by~\citet{buccinca2021trust} called out that when interacting with AI decision support, there is tension between cognitive engagement, which can lead to better decision outcomes, and subjective user experience, which may be influenced by ease of use and efficiency. Participants' comments suggest that there may exist such tension for the explanation types we studied. While example-based explanations encouraged more independent and deeper reasoning than feature-based explanations, these explanations could also be subjectively perceived as more cognitively demanding and time-consuming.

\subsubsection{Participant suggestions for improvement}

We asked how the decision support could have been improved. Here we summarize suggestions from the 20 participants who offered them.

\paragraph{Better support reasoning about example-based explanations} Following the above observations that example-based explanations were sometimes perceived to be less intuitive to use, participants suggested ways to improve them. As discussed, in pathway 2 reasoning, participants looked for evidence that the AI prediction might be wrong in the explanation based on an instance's similarity with examples or the impact of different feature values. To better support this, multiple participants suggested interactive example-based explanations that allow users ``\textit{to interactively change some of the factors in the person to be judged to see if that changed the prediction of the AI}'' (P24)  or more sophisticated sampling methods that either select contrasting examples with different ground truth labels or show examples that vary in different features to understand the model's ``\textit{decision boundaries}'' (P23). Inspired by feature-based explanations, some participants suggested combining the two explanation types by ``\textit{highlighting the keywords in examples}'' (P14), ``\textit{highlighting the different and similar features}'' (P20) between the instance and examples, or explicitly showing the impact of different feature values.

\paragraph{Better support intuition about AI limitations} Participants suggested several approaches that relate to improving their intuition about the AI system's limitations and boosting pathway 3 reasoning. In addition to suggesting that the AI system support include confidence or uncertainty quantification for each prediction, participants asked for support to have a global understanding of how the model assigns weights to different features in general (i.e., a global explanation~\cite{guidotti2018survey,gilpin2018explaining}) ``\textit{to show the influence of the attributes on the entire data set}'' (P20) or on what kind of cases the model is likely to be unreliable. Some noted that this kind of global model intuition can also help them better reason about local explanations. For example, P20 commented that if they were aware that the model is generally biased, they would discount the impact of demographic features when looking at the explanations. For users with more ML knowledge, P27 suggested providing more technical information about the underlying algorithm, training data, and how the explanations are generated: ``\textit{Are we using a deep learning model or are we using something more classic? Overall, where does the training data come from? ... What is the highlighting [method]? Is it SHAP? Is it LIME?}''

\paragraph{Better align explanations with human reasoning} Participants further pointed out that the current explanations do not fully align with human reasoning. For example, applying explanations to  embeddings at the token level (i.e., individual words)
does not align with features that people use to judge a biography. As discussed in Section~\ref{sec:rq2},  people tended to reason about features that were either ``\textit{two words together or phrases}'' (P25) or higher-level semantic features, linguistic markers, or writing styles. We note that recent XAI research has begun to explore how to ``translate'' raw features used by a model into higher-level concepts that are familiar to people~\cite{kim2017interpretability,ehsan2019automated}. Other suggestions relate to presenting the explanations as a human would, i.e., in a more interactive, conversational, and narrative manner rather than using bar charts. P2 said ``\textit{I saw the AI as more of a person or a friend explaining it to me. If the information was presented in a story format or a paragraph format, that would also help weigh my decision}.'' 

\section{Discussion}

We investigated the types of human intuition present in human-AI decision-making with explanations. We identified three intuition-driven pathways to override AI predictions: (1) using strong outcome intuition to disagree with the AI prediction; (2) applying intuition about features to reason about explanations to discredit the AI prediction; and (3) recognizing AI limitations through signals of prediction unreliability. 
In this section, we first discuss how these pathways can
support a more generalizable understanding of when and what explanations can help decision-making. Then we make design recommendations for AI support that helps decision-makers appropriately apply their own intuition.

\subsection{Understanding Overreliance in the Human-AI Decision-Making Process}
\label{sec:explainpriorwork}

To facilitate the responsible design of AI systems by preventing harmful overreliance~\cite{liao2022designing}, a fundamental understanding of people's decision-making process with AI is key. We believe the set of intuition-driven pathways we identified can be a useful tool to help understand why---and even anticipate when---inappropriate reliance may happen. We illustrate this with two use cases below.

\paragraph{These pathways can help explain the effect of different explanations} First, we can use the pathways to interpret the mixed results of prior empirical studies on different forms of example-based explanations. Contrary to our results, a previous study by~\citet{wang2021explanations} found no evidence that the example-based explanations they used can improve decisions over feature-based explanations. However, they adopted a different strategy to select examples, focusing only on examples where the model predictions are \textit{correct}, and always displaying one example for which the AI system predicts the same class and one example for which it predicts a different class. This approach may not provide easily recognizable prediction unreliability signals (pathway 3). We also suspect that this selection method may not result in examples that are consistently similar, then participants are less likely to utilize pathway 2 to find evidence that a prediction might be wrong. In another study,~\citet{kim2022hive} tested example-based explanations that showed representative prototypes for each class, rather than nearest neighbors of the given instance, and found less overreliance on incorrect AI predictions compared to feature-based explanations. We note that prototype-based explanations do not directly support pathway 3, but rather support pathway 2 when users recognize dissimilarity between the prototypes and the instance, which ~\citet{kim2022hive} also explicitly asked participants to rate.

\paragraph{These pathways can help understand which individuals will benefit more from AI support and why}  Understanding how individual differences affect the viability of each pathway can make it  
possible to anticipate who will or will not benefit from certain AI support. We discuss two concrete examples. The first is individuals with a low level of domain knowledge, where the only viable pathway is pathway 3 because pathways 1 and 2 require substantial domain knowledge. Since example-based explanations better support appropriate non-reliance through pathway 3, we expect it to be particularly helpful for people with low domain knowledge. Indeed, by using participants' decision accuracy over all instances in phase 2 (N=16) as a proxy measure of their domain knowledge, we found that the 5 participants with the lowest domain knowledge (average phase 2 accuracy = 43.8\%)
improved significantly with example-based explanations (average phase 3 accuracy = 70.0\%) but not with feature-based explanations (average phase 3 accuracy = 52.5\%). 

Second, we can anticipate that people whose own outcome intuition is highly consistent with the AI system will benefit less from AI support because of a weak pathway 1, which has been alluded to by prior works as complementary knowledge~\cite{bansal2019beyond,zhang2020effect}.
By using agreement between the participants' phase 2 decisions and AI predictions over all instances in phase 2 (N=16) as a proxy measure of \textit{complementary outcome intuition}, 
we find a moderate negative correlation with their phase 3 accuracy 
when using feature-based explanations ($r=-0.34$ for income prediction and $r=-0.41$ for biography classification).
Interestingly, there is little to no correlation between this proxy measure and accuracy with example-based explanations, possibly because the example-based explanations we used tend to encourage participants to utilize pathway 3,  
as discussed in Section~\ref{sec:rq2}.

\subsection{Design Recommendations for AI Decision-Support Systems}

We make the following recommendations for future work to develop more effective AI support that accounts for different types of human decision intuition as well as individual differences in them.

\paragraph{Accommodate human decision intuition of varied strengths} 
Our study suggests that even for the same decision task, human decision-makers may have a strong intuition for some instances but not others, presenting different opportunities for AI support. 
On the one hand, when people do not have a strong intuition, it would be difficult to identify when the AI system is incorrect through pathway 1 (but possible through other pathways).
On the other hand, when people have strong intuition, even though they may be more capable of overseeing the AI system, they may also be less motivated to engage with complex information. In this case, AI support should also avoid disrupting people's natural intuition and instead encourage people to examine their own intuition even when they are in agreement with the AI system. We can also envision adaptive or personalized systems that provide different kinds of information support (e.g., when to show a certain type of explanation) depending on the decision-maker's confidence in their own decision.

\paragraph{Make explanations compatible with human decision rationale} We observed that when using feature-based explanations, participants sometimes mistakenly override the AI prediction because they disagree with certain, potentially trivial, aspects of the explanation. 
Social science literature suggests that natural human explanations are predominantly \textit{qualitative}~\cite{miller2019explanation}---while people may have an intuition about which feature is relevant to the decision, or which feature is more relevant than another, they may not be able to articulate a precise quantification. This literature suggests that the common way to present feature-based explanations, which shows a contribution score for each feature, does not align with how people reason. Future work should explore more natural explanations that, for example, start with qualitative narratives and provide more precise information upon request. 
More broadly, our study identifies types of feature-based intuition that participants applied to reason about explanations (Section~\ref{result:common-themes}).
We hope they inspire future work to develop AI systems that leverage these reasoning patterns to design more compatible explanations.

\paragraph{Design explanations to facilitate people's use of intuition-driven pathways}  Previous studies have attributed the cause of overreliance on AI to a lack of cognitive engagement~\cite{liao2021human,gajos2022people,buccinca2021trust}. However, we show that even when people are relatively engaged (by thinking aloud), current XAI techniques do not reduce overreliance. This points to the inherent limitations of feature-based explanations in supporting participants to override incorrect predictions (i.e., the disruption to and incompatibility with natural intuition). 
To override AI predictions through pathway 2, we find people naturally look for evidence in explanations to discredit AI predictions (e.g., looking for multiple similar examples with a different or incorrect prediction in example-based explanations). 
We encourage future work to first explore what people consider to be evidence to discredit predictions for different types of explanations, and then to design interfaces to help people identify such evidence. 

\paragraph{Develop decision-support features that effectively reflect AI limitations} Lastly, we highlight the need for explanations to help people form intuition about AI limitations (pathway 3). 
Participants' interview comments suggested that global explanations and documentation of the model's failure cases can inform intuition about model limitations. 
While prior work has suggested that showing prediction uncertainty or confidence measures may be more effective than explanations for this purpose~\cite{zhang2020effect}, the former can suffer from miscalibration (i.e., the presented value does not correspond to the actual error probability).  
We also suggest further work on evaluation metrics that capture an XAI method's ability to \textit{reflect and communicate prediction unreliability}.  Similar metrics have been used to evaluate the calibration of uncertainty quantification~\cite{bhatt2021uncertainty,brier1950verification,naeini2015obtaining}---the estimated uncertainty should reflect the observed error rate. Our study suggests that such  metrics should also consider how people actually \textit{perceive} the information. 
For example, we can envision a metric that asks targeted users to rate the reasonableness of explanations and then quantifies the correlation with errors in the test data.

\subsection{Limitations}
\label{sec:limitation}
We acknowledge several inherent limitations of using a think-aloud method. First, by asking participants to verbalize their thought processes, our protocol may not completely resemble a realistic decision-making setting. It is possible that when not thinking aloud, especially in a low-stakes setting, people will be less cognitively engaged with AI explanations~\cite{gajos2022people} and some effects found in the study will not be observed. With this limitation, we again encourage readers to interpret the quantitative results with caution and focus on the trends. Second, think-aloud data are observational and do not allow for isolating the effect of different types of intuition or for identifying precise causes of how a participant arrived at a certain decision.  Therefore, our results focused on the themes that emerge from the observational data rather than attempting to draw conclusions about their precise relations and impact on decisions.

We also acknowledge the trade-off of a two-phase study design that showed participants the same set of instances in both phases. While this design allowed us to analyze cases where the human-alone decisions agree or disagree with the AI predictions separately, this design may have  strengthened, in some cases, participants' prior intuition more than a realistic human-AI decision-making setting.  That being said, we do not foresee any \textit{type} of intuition to be contingent on the set-up of our study design.

Our study was also limited by the choice of decision tasks and explanation methods, as well as the relatively small sample size. The two tasks are relatively low stakes and do not require specialized domain knowledge. They are also not representative of all data and feature types. 
One method in our study is a popular post-hoc feature-based explanation that is known to be not completely faithful to the underlying prediction model. The potential noise that can be introduced into decision-making by unfaithful post-hoc explanations, as opposed to directly interpretable models, is an open question that should be explored in future work. 
Our participants were not experts on these tasks and, despite our effort to diversify, the sample was biased toward more highly educated and ML-experienced individuals than the general population. Therefore, we acknowledge that the types of intuition identified in this work may not be complete or fully generalizable. We encourage future work to further study the interplay between human intuition and different types of explanations across domains and populations to verify the themes identified in this study.

\section{Conclusion}
We designed a mixed-methods study to understand how human decision-makers reconcile their own intuition with AI predictions and explanations. In light of prior work, which finds that feature-based explanations can increase overreliance on incorrect AI predictions, we study how decision-makers can apply their own intuition to override AI and have appropriate reliance.  
Our analysis of participants' think-aloud data  revealed three intuition-driven pathways to reduce overreliance on AI: (1) using strong outcome intuition to disagree with the AI prediction; (2) applying intuition about features to reason about explanations to discredit the AI prediction; and (3) recognizing AI limitations through signals of prediction unreliability. 
We use these pathways to explain why the example-based explanations we used helped lead to complementary human-AI performance and better supported appropriate reliance in comparison to feature-based explanations: they were less disruptive of people's natural intuition about outcomes, they better promoted inductive reasoning about features and the decision task generally, and, in particular, they provided strong and accurate signals of prediction unreliability.
These pathways also highlight the limitations of feature-based explanations, providing reasons why they lead to overreliance when the AI system is incorrect: they disrupt outcome intuition, may conflict with intuition about features, and do not present clear ways to reason about prediction unreliability.
Our findings provide fundamental knowledge about the human-AI decision-making process that could support a generalizable understanding of when and what explanations can help decision-making, and point to user needs for AI decision-support systems that better accommodate human decision intuition, are more compatible with human intuition, and support a more critical understanding of AI.

\section*{Acknowledgments}
This research was conducted at Microsoft Research. We thank our participants for their time and the reviewers for their feedback. We also thank Zana Bu\c{c}inca, Han Liu, Andreas Madsen, Stephanie Milani, and researchers in the Microsoft Research FATE group for their thoughtful suggestions.

\bibliographystyle{ACM-Reference-Format}
\bibliography{sample-base}

\appendix
\newpage

\section{Participant Info}
\label{sec:participanttable}

Table~\ref{tab:participants} contains demographic information about the study participants.

\begin{table}[h!]
\caption{Table of participant information.  For ML and XAI knowledge, participants rated themselves using the following scale: 0 = ``None''; 1 = ``Limited experience, I know the basic concepts''; 2 = ``I have experienced them often or they are part of my day-to-day life''; 3 = ``I consider myself an expert on them.''}
\resizebox{\textwidth}{!}{\begin{tabular}{@{}lllccl@{}}
\toprule
    & \multicolumn{1}{c}{\textbf{Education}} & \multicolumn{1}{c}{\textbf{Role}} & \textbf{\begin{tabular}[c]{@{}c@{}}ML \\ Knowledge\end{tabular}} & \textbf{\begin{tabular}[c]{@{}c@{}}XAI \\ Knowledge\end{tabular}} & \multicolumn{1}{c}{\textbf{\begin{tabular}[c]{@{}c@{}}Experimental\\ Condition\end{tabular}}} \\ \midrule
P1  & Graduate Degree                        & Software Engineer                 & 2                                                                & 0                                                                 & Income Prediction                                                                             \\
P2  & College                                & Software Engineer                 & 0                                                                & 0                                                                 & Income Prediction                                                                             \\ 
P4  & Graduate Degree                        & Software Engineer                 & 3                                                                & 2                                                                 & Income Prediction                                                                             \\
P8  & College                                & Data Scientist                    & 2                                                                & 2                                                                 & Biography Classification                                                                            \\
P9  & Graduate Degree                        & Applied Scientist                 & 3                                                                & 2                                                                 & Income Prediction                                                                          \\
P11 & Graduate Degree                        & PhD Student                       & 3                                                                & 2                                                                 & Income Prediction                                                                             \\
P12 & College                                & Engineer                          & 3                                                                & 3                                                                 & Income Prediction                                                                             \\
P14 & Graduate Degree                        & PhD Student                       & 2                                                                & 2                                                                 & Biography Classification                                                                            \\
P15 & Graduate Degree                        & PhD Student                       & 3                                                                & 2                                                                 & Biography Classification                                                                            \\
P16 & College                                & Editor / Data analyst               & 1                                                                & 1                                                                 & Biography Classification                                                                            \\
P17 & Graduate Degree                        & PhD Student                       & 2                                                                & 3                                                                 & Income Prediction                                                                             \\
P18 & Graduate Degree                        & Software Engineer                 & 1                                                                & 1                                                                 & Biography Classification                                                                            \\
P19 & Graduate Degree                        & Professor                         & 2                                                                & 2                                                                 & Biography Classification                                                                            \\
P20 & Graduate Degree                        & Research Engineer / Scientist     & 2                                                                & 2                                                                 & Income Prediction                                                                             \\
P21 & Graduate Degree                        & Engineer                          & 3                                                                & 2                                                                 & Biography Classification                                                                            \\
P22 & College                                & Software Engineer                 & 1                                                                & 0                                                                 & Biography Classification                                                                            \\
P23 & Graduate Degree                        & PhD Student                       & 2                                                                & 2                                                                 & Income Prediction                                                                             \\
P24 & Graduate Degree                        & PhD Student                       & 3                                                                & 3                                                                 & Income Prediction                                                                             \\
P25 & Graduate Degree                        & PhD Student                       & 3                                                                & 3                                                                 & Biography Classification                                                                            \\
P26 & Graduate Degree                        & PhD Student                       & 2                                                                & 2                                                                 & Income Prediction                                                                             \\
P27 & College                                & PhD Student                       & 3                                                                & 2                                                                 & Biography Classification                                                                            \\
P28 & Graduate Degree                        & Software Engineer                 & 1                                                                & 0                                                                 & Income Prediction                                                                             \\
P30 & Graduate Degree                        & Research Engineer / Scientist     & 3                                                                & 2                                                                 & Biography Classification                                                                            \\
P31 & Graduate Degree                        & Data Scientist                    & 3                                                                & 2                                                                 & Biography Classification                                                                            \\
P32 & College                                & PhD Student                       & 1                                                                & 1                                                                 & Income Prediction                                                                             \\
P33 & Graduate Degree                        & PhD Student                       & 1                                                                & 1                                                                 & Biography Classification                                                                            \\
\bottomrule
\end{tabular}}
\label{tab:participants}
\end{table}

\section{Post-Study Interview Questions}\label{appdx:interview}

The following questions were asked in the post-study interviews.

\begin{itemize}
    \item Can you describe your general strategies to make decisions when AI was not
    available in phase 2?
    \item What difference did introducing the AI in phase 3 make? 
    \item How did you reason with AI prediction and explanations? 
    \item What strategies did you take in phase 3 when you had feature contributions versus similar examples? Which explanation type did you prefer?
    \item In general, do you think you are better at this task than the AI? If yes, would you still want to use the AI if you were asked to perform the task again?
    \item Is there something else besides feature contribution or similar examples that you wish to know or that would have helped you make better decisions?
\end{itemize}

\section{User Study Instructions}\label{appdx:instructions}

\begin{figure}[h!]
     \centering
     \includegraphics[width=0.95\textwidth]{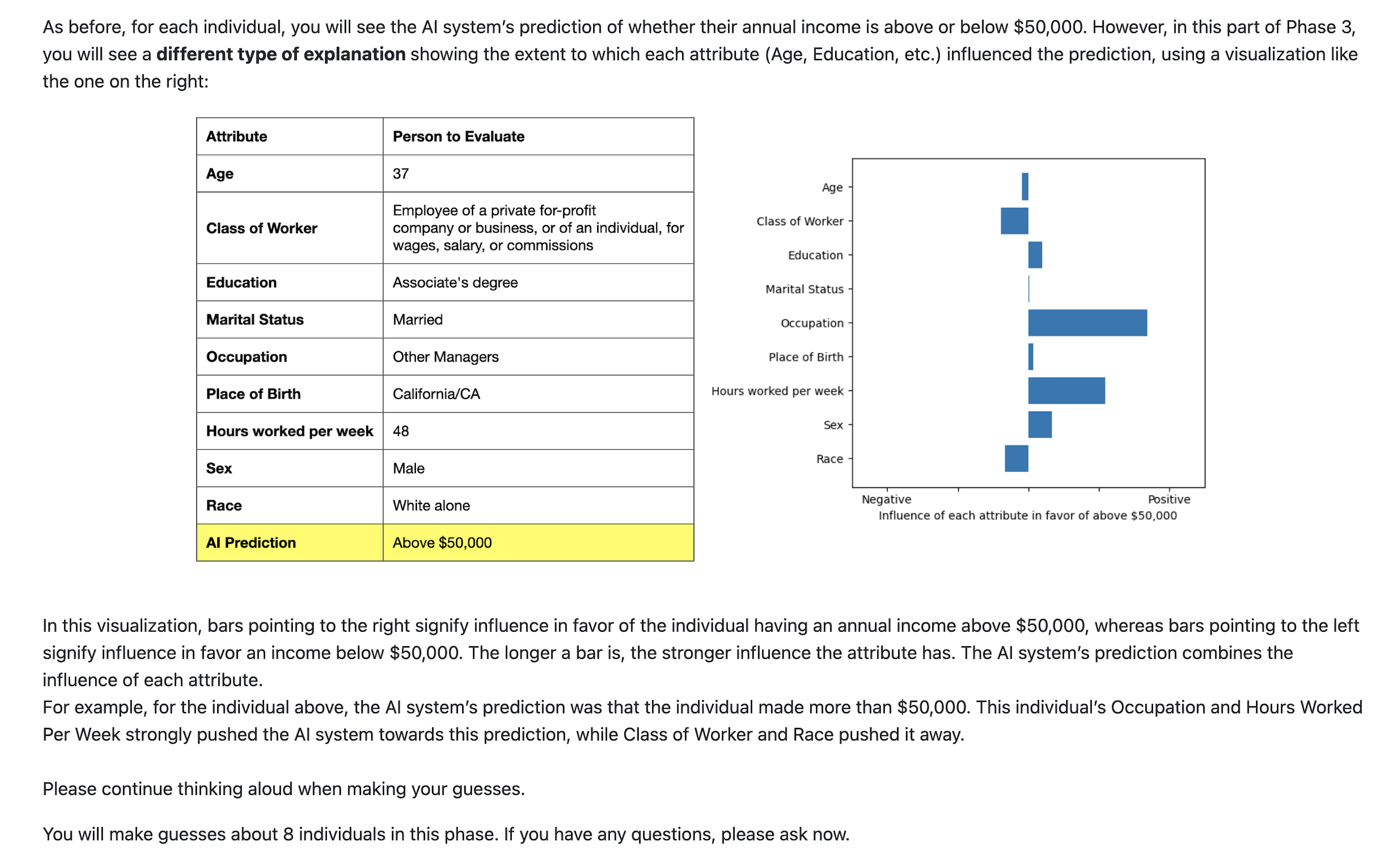}
     \caption{Interface and instructions for how to use feature-based explanations in the income prediction task.}
     \label{fig:interface_tab_F}
\end{figure}

\begin{figure}[h!]
     \centering
     \includegraphics[width=0.95\textwidth]{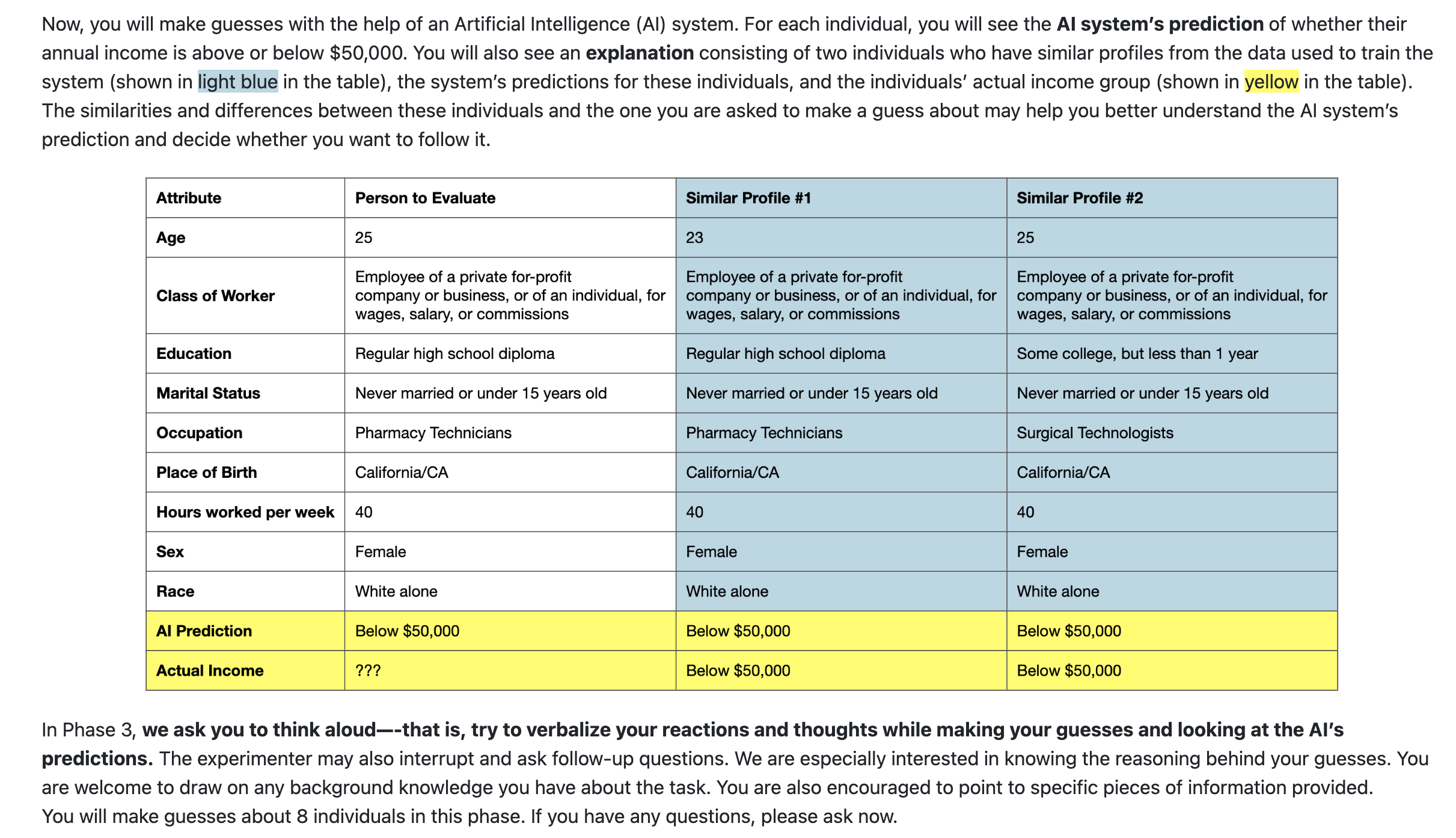}
     \caption{Interface and instructions for how to use example-based explanations in the income prediction task.}
     \label{fig:interface_tab_E}
\end{figure}

\begin{figure}[h!]
     \centering
     \includegraphics[width=0.95\textwidth]{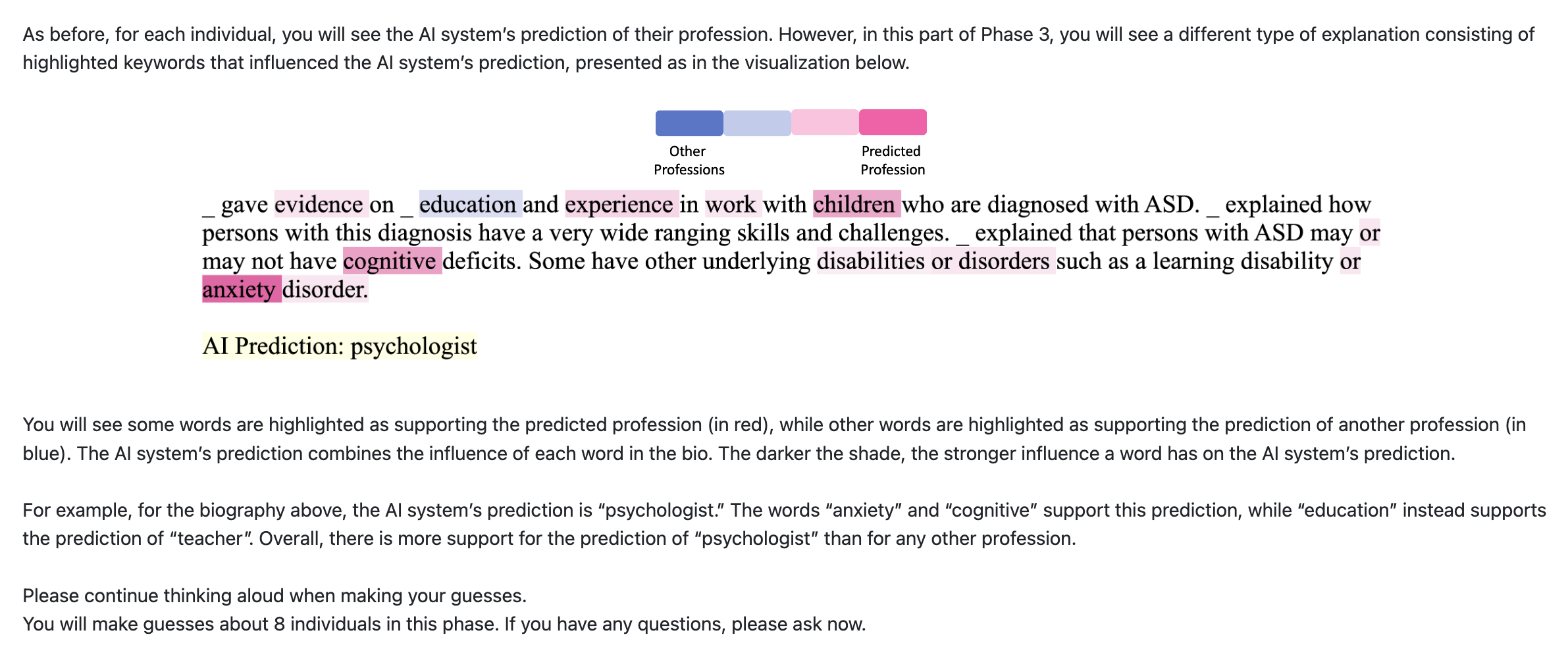}
     \caption{Interface and instructions for how to use feature-based explanations in the biography classification task.}
     \label{fig:interface_text_F}
\end{figure}

\begin{figure}[h!]
     \centering
     \includegraphics[width=0.95\textwidth]{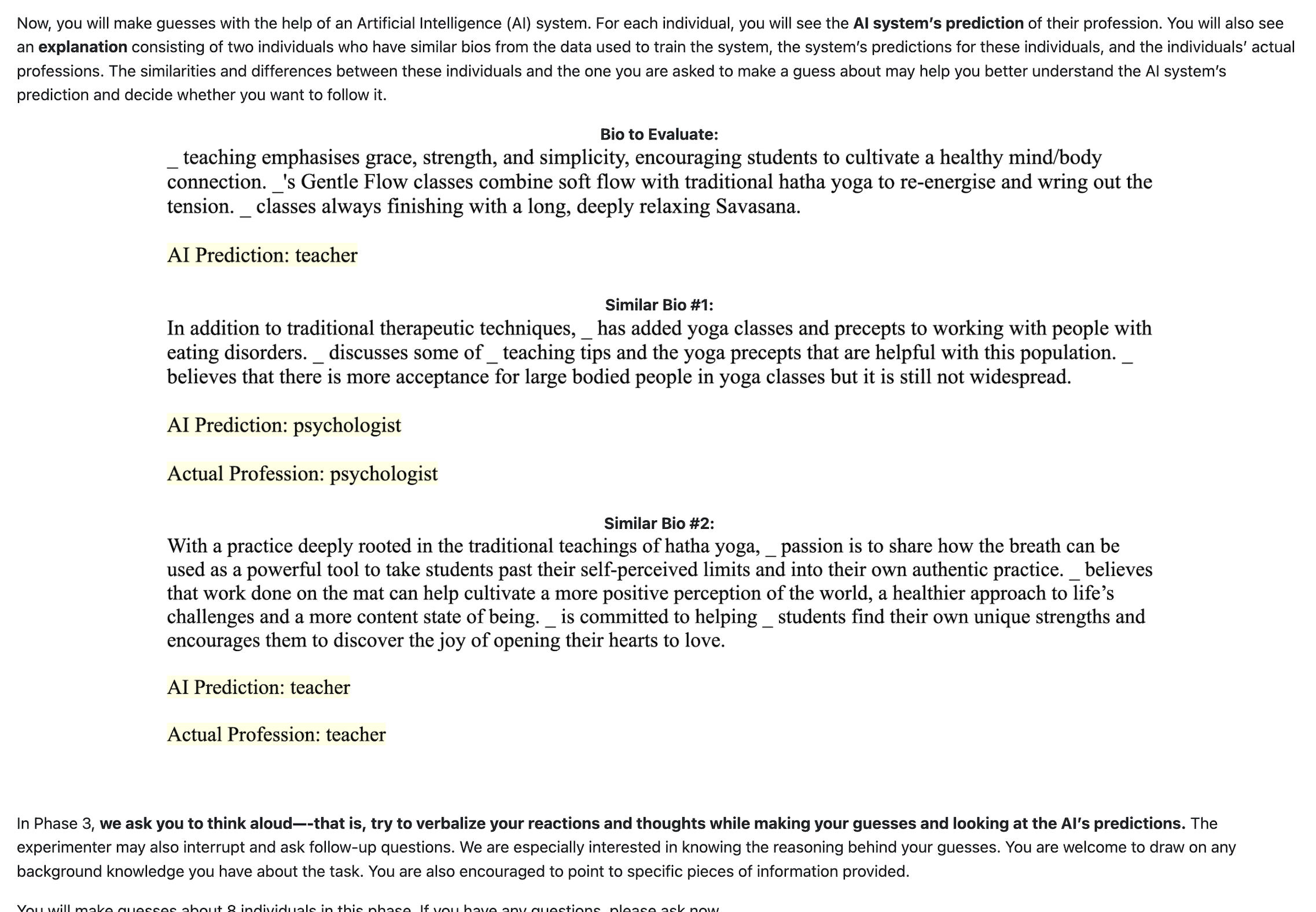}
     \caption{Interface and instructions for how to use example-based explanations in the biography classification task.}
     \label{fig:interface_text_E}
\end{figure}

Figures~\ref{fig:interface_tab_F}--\ref{fig:interface_text_E} show the instructions that study participants were given. These variants correspond to cases in which the participant saw example-based explanations before feature-based explanations.

\end{document}